\documentclass{aa}  
\usepackage{etex}
\reserveinserts{18}
\usepackage{morefloats}
\usepackage{color,soul}
\usepackage{graphicx}
\usepackage{txfonts}
\usepackage{caption}
\usepackage{subcaption}
\usepackage[normalem]{ulem}
\begin{document}

   \title{Onset of 2D magnetic reconnection in the solar photosphere, chromosphere and corona}

   \author{B. Snow
          \inst{1,2}
          \and
          G. J. J. Botha
          \inst{1}
          \and
          J. A. McLaughlin
          \inst{1}
          \and
          A. Hillier
          \inst{3}
          }

   \institute{Northumbria University, Newcastle upon Tyne, NE1 8ST, UK\\
   \and
   University of Sheffield, Sheffield, S3 7RH, UK \\
              \email{b.j.snow@sheffield.ac.uk}
         \and
             {CEMPS, University of Exeter, Exeter, EX4 4QF, UK}\\
             }

  \date{Received; Accepted}

  \abstract
{}
{We aim to investigate the onset of 2D time-dependent magnetic reconnection that is triggered using an external (non-local) velocity driver located away from, and perpendicular to, an equilibrium Harris current sheet. Previous studies have typically utilised an internal trigger to initiate reconnection, for example initial conditions centred on the current sheet. Here, an external driver allows for a more naturalistic trigger as well as the study of the earlier stages of the reconnection start-up process.}
{Numerical simulations solving the compressible, resistive magnetohydrodynamics (MHD) equations were performed to investigate the reconnection onset within different atmospheric layers of the Sun, namely the corona, chromosphere and photosphere.}
{A reconnecting state is reached for all atmospheric heights considered, with the dominant physics being highly dependent on atmospheric conditions. The coronal case achieves a sharp rise in electric field (indicative of reconnection) for a range of velocity drivers. For the chromosphere, we find a larger velocity amplitude is required to trigger reconnection (compared to the corona). For the photospheric environment, the electric field is highly dependent on the inflow speed; a sharp increase in electric field is obtained only as the velocity entering the reconnection region approaches the Alfvén speed. Additionally, the role of ambipolar diffusion is investigated for the chromospheric case and we find that the ambipolar diffusion alters the structure of the current density in the inflow region.}
{The rate at which flux enters the reconnection region is controlled by the inflow velocity. This determines all aspects of the reconnection start-up process, that is, the early onset of reconnection is dominated by the advection term in Ohm’s law in all atmospheric layers. A lower plasma-$\beta$ enhances reconnection and creates a large change in the electric field. A high plasma-$\beta$ hinders the reconnection, yielding a sharp rise in the electric field only when the velocity flowing into the reconnection region approaches the local Alfvén speed.}



\maketitle

\section{Introduction}

Magnetic reconnection is a topological restructuring of a magnetic field causing a change in connectivity of its fieldlines \citep{Priest2000}. In this process, the magnetic energy is converted into kinetic energy and thermal energy of the plasma. Reconnection plays an important role in many dynamical phenomena in the solar atmosphere including the photosphere, for example Ellerman bombs \citep[e.g.][]{Ellerman1917,Reid2016}, chromosphere, for example penumbral microjets \citep[e.g.][]{Katsukawa2007} and calcium jets \citep[e.g.][]{Shibata2007,Morita2010}, and the corona, for example x-ray jets \citep[e.g.][]{Shibata1992, Savcheva2007} and flares \citep[e.g.][]{Moore2001}. Comprehensive reviews of magnetic reconnection can be found in \cite{Zweibel2009}, \cite{Yamada2010} and \cite{Pontin2012}. 

There are several models for steady-state reconnection which can be separated broadly into two categories: slow and fast. The model of Sweet-Parker reconnection \citep{Parker1957,Sweet1958} has a long diffusion region between oppositely directed magnetic fieldlines. The rate at which the magnetic fields diffuse predicts an Alfv\'en Mach number of the inflow region, $M_A$, equal to the inverse root of the Lundquist number, that is $M_{A} = S^{-1/2}$. In astrophysical plasmas, the Lundquist number can be many orders of magnitude larger than unity, resulting in a very small Alfv\'en Mach number, hence the name slow reconnection. Slow reconnection is not sufficient to explain observed phenomena, e.g. the energy release timescales necessary for solar and stellar flares \citep[][p. 410]{Aschwanden2005}. Amendments have been made on the Sweet-Parker model to create fast reconnection models, for example  \cite{Petschek1964} and flux pile-up models \citep{Priest1986}. The Petschek model achieves fast reconnection by including two pairs of slow-mode shocks which act to carry flow away from the diffusion region. This results in a diffusion region that shortens as the inflow rate increases. Flux pile-up models have a rise in magnetic field energy entering the diffusion region and a slow-mode expansion. In the flux pile-up models, the diffusion region becomes long and thin, and the reconnection rate can exceed the maximum Petschek reconnection rate \citep{Priest1986}. These models all relate to steady-state reconnection, whereas in the solar atmosphere reconnection is time dependent.

The time evolution of a reconnection event has been studied via the incompressible Taylor problem, where a small boundary perturbation is applied to a stable slab plasma, as reviewed in \cite{Bhattacharjee2004}. This is typically referred to as forced or driven reconnection. The linear and non-linear phases of this process can be studied separately to analyse the evolution of a reconnection event, and the formation of current sheets and magnetic islands \citep{Wang1992}. Numerical simulations of the Taylor problem use an external perturbation to develop a localised current \citep[][]{Fitzpatrick2003b} yielding results consistent with analytical solutions \citep[][]{Hahm1985,Dewar2013}. The non-linear phase is only weakly dependant on the resistivity \citep{Wang1996}. The external velocity perturbation reflects in the enclosed domain producing narrow spikes in the current \citep[][]{Fitzpatrick2003}.

When investigating reconnection in the solar atmosphere it is common to study the initialisation and development of magnetic reconnection through an internal trigger, that is various perturbations specified as initial conditions that are centred on the current sheet.
These include invoking locally-enhanced resistivity or small velocity perturbations inside the current sheet \citep[e.g.][]{Ugai1977,Arber2006,Malyshkin2010}. However, such an initialisation may not always be applicable for reconnection in the highly dynamic solar atmosphere.


Stable current sheets can form in a number of ways throughout the solar atmosphere \citep[][Chapter 2]{Priest2000}. For example, current sheets can form around rational surfaces of a force-free magnetic field, through collapse of magnetic structures, and by compression of the plasma. Analytically, the induction equation can be solved to find a stable current sheet with velocity inflow \citep[][p. 94]{Priest2000}.

It is known that waves and flows are ubiquitous in the solar atmosphere (e.g. \citealt{1983SoPh...88..179E,2005LRSP....2....3N,2005RSPTA.363.2743D,2007AdSpR..39.1804N,2007Sci...317.1192T}). Sub-Alfv\'enic flows are observed at coronal \citep[e.g.][]{2005LRSP....2....3N}, chromospheric \citep[e.g.][]{Jess2015} and photospheric \citep[e.g.][p.19]{priest1984solar} atmospheric heights, and flux cancellation has been observed at photospheric levels \citep[e.g.][]{Nelson2016}. Thus, these waves and flows can, at some time, encounter a current sheet and this can initiate reconnection. However, the process by which reconnection can then develop as a result of this external driver is still to be studied in detail. 

In the solar chromosphere, the temperature and density allow both neutral and ionised particles to exist, hence the plasma is partially ionised. This results in Cowling resistivity that acts perpendicular to magnetic field, in addition to Spitzer resistivity. Previous work investigating the implications of partial ionisation includes studies of flux emergence \citep{Leake2006, Arber2007}, wave dissipation \citep{Leake2005}, tearing mode instabilities \citep{Leake2012}, chromospheric current sheet collapse \citep{Arber2009}, and the evolution of slow mode shocks \citep{Hillier2016}. For partially-ionised coalescing loops it has been shown that the ionisation increases the amount of reconnected magnetic flux, but the reconnection rate remains unchanged \citep{Smith2008}. The effects of partial ionisation can also be used to explain the existence of penumbral microjets \citep{Sakai2008}. For strong chromospheric magnetic field strengths, the length scale of a tearing mode instability can become comparable to the kinetic length scales that have been hypothesised to be necessary for fast reconnection \citep{Singh2015}. For chromospheric reconnection the Hall term, $\textbf{J} \times \textbf{B}$, in Ohm's law can generally be neglected \citep[][]{Malyshkin2011}. A review of the effects of partial ionisation can be found in \cite{Zweibel2011}.

This paper describes an investigation into the onset of 2D time-dependent magnetic reconnection that is triggered using an (external or non-local) sub-Alfv\'enic velocity driver specified perpendicular to an equilibrium Harris current sheet, removed from the centre of the domain. The physical motivation of this model is to study how reconnection events can be triggered by converging flows of the dynamic solar atmosphere itself. Reconnection start-up and development is investigated in coronal (fully-ionised), chromospheric (partially-ionised) and photospheric (weakly-ionised) atmospheric conditions. In order to understand the differences between coronal, chromospheric  and photospheric reconnection, the coronal model is first presented and analysed as a reference model, and then a parameter study is performed. This allows the effects of each parameter to be studied independently.
We note that this paper is studying the time-dependent onset of reconnection, and should not be confused with  steady-state reconnection, which has been well studied from both an analytical and numerical description, for example \citep{Priest2000}.

The structure of this paper is as follows. First the computational model is described (Section \ref{sec_sim}). Then a reference coronal case is investigated (Section \ref{sec_fid}). A parameter study is then formed on this reference model, investigating the velocity dependence (Section \ref{sec_vel}) and plasma-$\beta$ (Section \ref{sec_beta}). The role of ambipolar diffusion is investigated in Section \ref{sec_ambi} for a chromospheric case. Finally, comparisons are made regarding the onset of reconnection in the photosphere, chromosphere and corona (Section \ref{sec_comp}).

\begin{figure}[!h]
\begin{center}
\setlength{\unitlength}{0.4cm}
\begin{picture}(16,16)
\put(0,0){\line(0,16){16}} 
\put(0,16){\line(16,0){16}} 
\put(16,16){\line(0,-16){16}} 
\put(0,0){\line(16,0){16}} 
\multiput(0,8)(0.495,0){33}{\line(1,0){0.2}} 
\multiput(4,6)(0,0.5){9}{\line(0,1){0.2}} 
\multiput(12,6)(0,0.5){9}{\line(0,1){0.2}} 
\put(12.5,8.5){$2\delta$} 
\multiput(4,6)(0.485,0){17}{\line(1,0){0.2}} 
\multiput(4,10.2)(0.485,0){17}{\line(1,0){0.2}} 
\put(9.5,10.7){$2L$} 
\put(0,4){\line(16,0){16}} 
\multiput(0,0)(1,0){16}{\line(1,1){1}} 
\multiput(0,1)(1,0){16}{\line(1,1){1}} 
\multiput(0,2)(1,0){16}{\line(1,1){1}} 
\multiput(0,3)(1,0){16}{\line(1,1){1}} 
\put(0,12){\line(16,0){16}} 
\multiput(0,12)(1,0){16}{\line(1,1){1}} 
\multiput(0,13)(1,0){16}{\line(1,1){1}} 
\multiput(0,14)(1,0){16}{\line(1,1){1}} 
\multiput(0,15)(1,0){16}{\line(1,1){1}} 
\put(8,12){\vector(0,-1){2}} 
\put(8.5,12){\vector(0,-1){1.5}} 
\put(9,12){\vector(0,-1){1}} 
\put(7.5,12){\vector(0,-1){1.5}} 
\put(7,12){\vector(0,-1){1}} 
\put(8,4){\vector(0,1){2}} 
\put(8.5,4){\vector(0,1){1.5}} 
\put(9,4){\vector(0,1){1}} 
\put(7.5,4){\vector(0,1){1.5}} 
\put(7,4){\vector(0,1){1}} 
\put(-1.4,0){$-8$} 
\put(-1.4,3.8){$-4$} 
\put(-0.8,7.7){$0$} 
\put(-0.8,11.8){$4$} 
\put(-0.8,15.5){$8$} 
\put(-1.8,7.7){\rotatebox{90}{$z$}}
\put(-0.8,-1){$-16$} 
\put(7.5,-1){$0$} 
\put(15.4,-1){$16$} 
\put(7.5,-2){$y$} 
\end{picture}
\vspace{0.5cm}
\end{center}
\caption{Computational domain. $y$ ranges between $\pm 16$, $z$ ranges between $\pm 8$. The drivers are defined for $|z| \geq 4$, denoted by the arrows. Shaded region indicates the damping zone. A reconnection region of width $2\delta$ and height $2L$ is fitted to the domain. The line $z=0$ is the location of the Harris current sheet. $y$ and $z$ are normalised by $10^4$ km.}
\label{figdomain}
\end{figure}
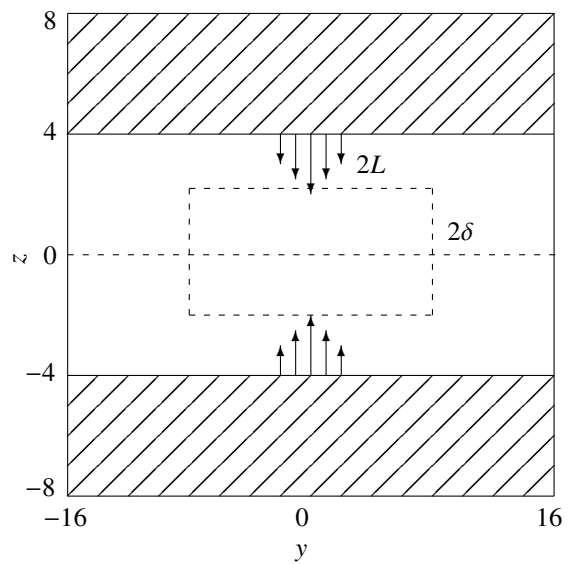

\section{Computational set-up and diagnostics} \label{sec_sim}

Simulations have been performed using Lare3D \citep{Arber2001} in the Cartesian $yz-$plane, with an invariant $x$-direction. Lare3D solves the compressible, resistive MHD equations given by
\begin{eqnarray}
\frac{\partial \rho}{\partial t} &=& - \nabla \cdot (\rho \textbf{v}), \label{eqnmhd1}\\
\frac{\partial \textbf{v}}{\partial t} + \textbf{v} \cdot \nabla \textbf{v} &=& \frac{1}{\rho} \textbf{J} \times \textbf{B} - \frac{1}{\rho} \nabla P,  \\
\frac{\partial \textbf{B}}{\partial t} &=& - \nabla \times \textbf{E},  \\
\frac{\partial \epsilon}{\partial t} + \textbf{v} \cdot \nabla \epsilon &=& - \frac{P}{\rho} \nabla \cdot \textbf{v} + \frac{\eta}{\rho} J^2  + \frac{\eta _\perp}{\rho} J_\perp ^2, \label{eqnieeqn}\\
\textbf{E} +\textbf{v} \times \textbf{B} &=& \eta \textbf{J} +\eta _{\perp} \textbf{J} _{\perp},  \label{eqnohmslaw}\\
\nabla \times \textbf{B} &=& \mu _{0} \textbf{J},  \\
P &=& \frac{\rho k_B T}{\mu _m}, \\
\epsilon &=& \frac{P}{\rho (\gamma -1)} +(1-\xi _n )\frac{\chi_{i}}{\bar{m}}, \label{eqninternal} \\
\eta _{\perp} &=& \frac{\xi _n ^2 B^2}{\alpha _n}, \label{eqneta}\\
\mu _m &=& \frac{\bar{m}}{2-\xi _n},  \label{eqnmhd9}
\end{eqnarray}
for density $\rho$, time $t$, velocity $\textbf{v}$, current density $\textbf{J}$, magnetic field $\textbf{B}$, internal energy $\epsilon$, Spitzer resistivity $\eta$, electric field $\textbf{E}$, pressure $P$, temperature $T$, neutral fraction $\xi _n$ and reduced mass $\mu _m$. $\textbf{J}_{\perp}$ is the current density perpendicular to the magnetic field. $\eta_{\perp}$ is an additional resistivity due to partial ionisation that acts on the perpendicular current density only. The universal constants are Boltzmann's constant $k_B = 1.381 \times 10^{-23}$ m$^2$ kg s$^{-2}$ K$^{-1}$, ionisation energy of hydrogen $\chi _i = 2.179 \times 10^{-18}$ J and permeability of free space $\mu _0 = 4 \pi \times 10^{-7}$ m kg s$^{-2}$ A$^{-2}$. Spitzer resistivity $\eta$ is uniform across the domain. Lare3D normalises parameters based on a normalisation density $\rho _0$, length $l_0$ and magnetic field strength $B_0$. In this paper, $l_0=10^4$ m, $B_0=0.002$ T and $\rho _0$ varies in the corona, photosphere and chromosphere. The numerical resistivity has been tested and is of the order $10^{-8}$ $\Omega$m (see Appendix \ref{appa}).

The neutral fraction $\xi _n$ is calculated using the modified Saha equation \citep{Thomas1961}. The neutral fraction is a function of density and temperature, that is $\xi _n = \xi _n (\rho ,T)$, as implemented in Lare3D by \cite{Leake2005}.

Our paper investigates time-dependent magnetic reconnection in a similar set-up as the Taylor problem, however we use careful treatment of the boundaries to prevent reflections of our initial perturbation (whereas the Taylor problem has reflecting boundaries). We apply this methodology using atmospheric conditions at various heights of the solar atmosphere allowing us to consider a naturalistic onset of reconnection occurring throughout the Sun.

A sketch of the computational domain is shown in Figure \ref{figdomain}, where the damping region is indicated by the shaded region. The initial magnetic field is specified using a Harris current sheet defined as 
\begin{equation}
B_y (z) = \tanh (z) 
\end{equation} 
where $B_y$ is the magnetic field component in the Cartesian $yz-$plane. $B_z =0$ initially. The maximum current is in the centre of the domain indicated by the dashed line at $z=0$ in Figure \ref{figdomain}.

For a uniform temperature the pressure balance equation is solved analytically to find the equilibrium density as 
\begin{equation}\label{den_dist}
\rho = \frac{1}{\beta} \cosh ^{-2} \left( z \right) +1.
\end{equation}

The physical extent of the domain is $320$ km  $\times 160$ km, with $512 \times 512$ cells and a grid size is $\Delta y=0.625$ km by $\Delta z =0.3125$ km. This domain size is sufficiently large in the $y$-direction that any shocks produced in the simulation do not reach the outflow boundaries. It also allows us to consider an isolated parameter regime in the $z$-direction, neglecting variations with altitude, i.e., the box size is smaller than the pressure scale height. The normalised grid dimensions are $-16 \le y \le 16$ and $-8 \le z \le 8$. A grid convergence test was performed using 256, 512 and 1024 cells and there are no significant quantitative or qualitative changes in the output for increased resolution, therefore $512 \times 512$ is sufficient.


\subsection{Velocity driver}

The reconnection is triggered by sub-Alfv\'enic velocity drivers specified far away from the centre of the current sheet at $4\leq z \leq8$ and $-8 \leq z \leq -4$. Both velocity drivers are of the form 
\begin{equation}
v_z(y)=A e^{-y^2 /3}. 
\end{equation}
This velocity propagates across the domain dragging magnetic field energy towards the centre of the domain and triggers the reconnection. 

When the velocity fronts meet, a fast-mode reflection occurs. It propagates towards the driven boundary and if left untreated bounces off the upper and lower boundaries and back into the reconnection region causing unwanted phenomena. To limit the influence of the reflection on the simulation, a damping region is specified between $-8\le z \le -4$ and $4 \le z \le 8$. One can damp out the unwanted perturbation by acknowledging the velocity in this region has two components, the driver $v_d$ and the perturbation $v_p$, i.e. $v=v_{d} +v_{p}$. Kinetic energy damping is applied to $v_{p}$ only and the driven velocity is applied uniformly between $-8 \leq z \leq -4$ and $4 \leq z \leq 8$. This has the effect of damping out the unwanted perturbations whilst maintaining the imposed driven velocity.

\subsection{Reconnection region}

A reconnection region of width $2 \delta$ and length $2 L$ is fitted to the domain, see Figure \ref{figdomain}. $\delta$ is estimated from the half width at half maximum (HWHM) of the current density $\textbf{J}$. $L$ is estimated by the point along the centre line $(z=0)$ at which the outflow velocity is maximum. This defines $L$ as the point at which the plasma ceases to be accelerated away from the reconnection region.

\subsection{Electric field}

The electric field $\textbf{E}$ is calculated using Ohm's law (Equation \ref{eqnohmslaw}).
In our computational domain there is only one non-zero component of $\textbf{E}$ that is perpendicular to the $yz$-plane, that is $\textbf{E}=(E_x,0,0)$. The electric field can be separated into a diffusion term $\eta \textbf{J} +\eta _\perp \textbf{J}_\perp$ and an advection term $\textbf{v} \times \textbf{B}$. We note that the only non-zero component of $\textbf{J}$ is $J_x$. This is equivalent to $J_\perp$ for our 2D simulation.

In 2D systems, reconnection can only occur at a null point. Therefore, in steady-state reconnection, the reconnection rate is given by the resistive electric field $\eta \textbf{J}$ at the null point. This also provides a good measure of reconnection rate during time-dependent simulations of reconnection, for example tearing-mode studies. 
In this paper, we are interested in the onset of reconnection, namely the transition from ideal to resistive dynamics. At early times of the simulation, the system is not reconnecting despite current being present. As the simulation progresses, there is a continuous change moving from the ideal phase to the resistive phase, and the flux inflow ($\textbf{v} \times \textbf{B}$ in the inflow region) does not equate to the flux transfer ($\eta \textbf{J}$ at the x-point). Therefore, using $\eta \textbf{J}$ is potentially misleading since there is an imbalance of the fluxes.
To capture the dynamics of the reconnection region, including the onset of reconnection, the maximum electric field in the diffusion region is used. This takes into account the motional electric field $\textbf{v} \times \textbf{B}$ in addition to the resistive electric field $\eta \textbf{J}$. By looking at the maximum electric field in the diffusion region, we gain insight into the dynamics in and around the reconnection site.
Note that as the reconnection evolves, the motional electric field $\textbf{v} \times \textbf{B}$ stagnates and the resistive electric field $\eta \textbf{J}$ increases. This results in the fluxes tending towards an equilibrium and means that towards the end of the simulation, the reconnection rate $\eta \textbf{J}$ becomes an appropriate metric since the system is tending towards a steady-state.

\section{Coronal (fully-ionised) environment}\label{sec_fid}

We first consider a  coronal-type atmosphere driven with a sub-Alfv\'enic velocity. The key parameters for this study are in Table \ref{tabbasepar1}.

\begin{table}[!h]
\caption{Reference parameters.}
     $$ 
         \begin{array}{p{0.5\linewidth}ll}
            \hline
            \noalign{\smallskip}
            Parameter [Units]   & \mbox{Symbol} &  \mbox{Value} \\
            \noalign{\smallskip}
            \hline
            \noalign{\smallskip}
            Plasma-$\beta$ & & 0.002     \\
            Length normalisation [m] & l_0 & 10^4 \\
            Magnetic normalisation [T] & B_0 & 0.002 \\
            Density normalisation [kg m$^{-3}$] & \rho _0 & 10^{-16} \\
            Temperature [K] & T & 10^6             \\
            Resistivity [$\Omega$m] & \eta & 10^{-6} \\
            Time normalisation [s] & \Delta t & 24  \\
            Alfv\'en Mach number & M_{A(driver)} & 0.003 \\
            Lundquist number & S & 2.5 \times 10^{10} \\
            \noalign{\smallskip}
            \hline
         \end{array}
     $$ 
\tablefoot{The Lundquist number above is the maximum value in the domain that occurs on the driven boundaries. In the reconnection region the maximum Lundquist number is $S < 10^{7}$.}
\label{tabbasepar1}
\end{table}

\begin{figure*}
\centering
\includegraphics[scale=0.04,clip=true, trim=2cm 2cm 1cm 3cm]{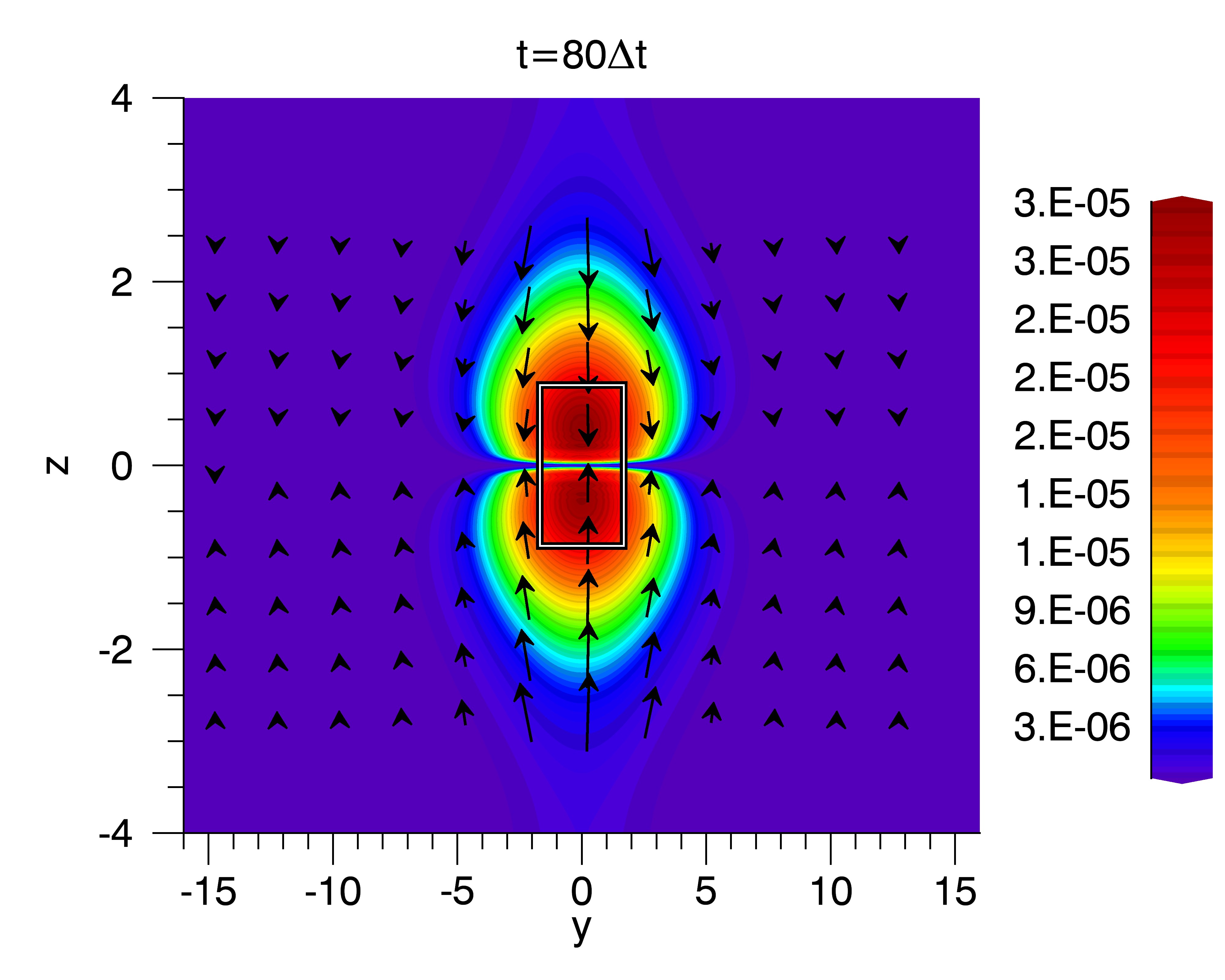}
\includegraphics[scale=0.04,clip=true, trim=2cm 2cm 1cm 3cm]{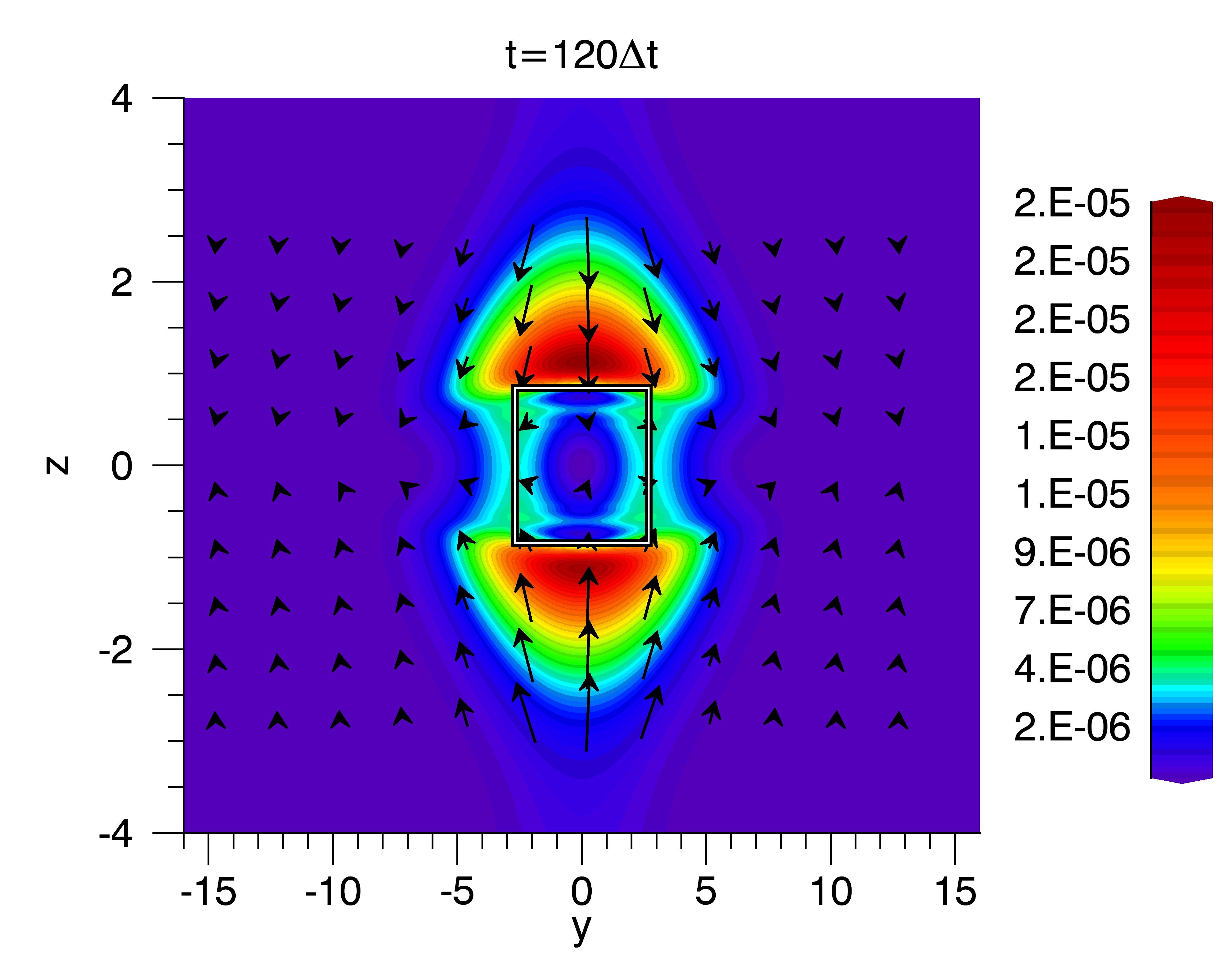}
\includegraphics[scale=0.04,clip=true, trim=2cm 2cm 1cm 3cm]{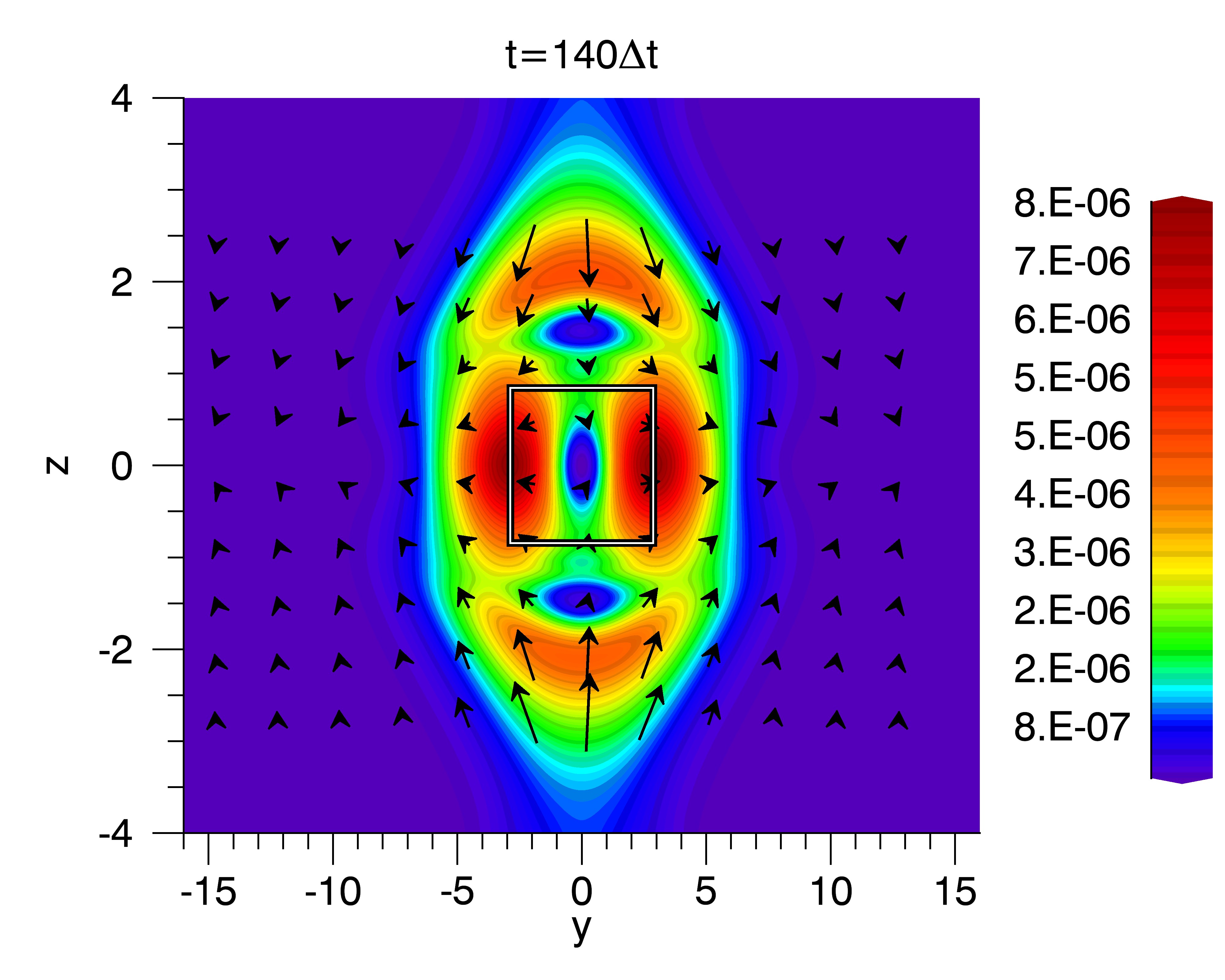}
\caption{Fast-mode reflection shown in kinetic energy at times 80$\Delta$t, 120$\Delta$t and 140$\Delta$t for the coronal atmosphere. The white box frame denotes the reconnection region.}
\label{figfullyref}
\end{figure*}

\begin{figure*}
\centering
\includegraphics[scale=0.04,clip=true, trim=2cm 2cm 1cm 3cm]{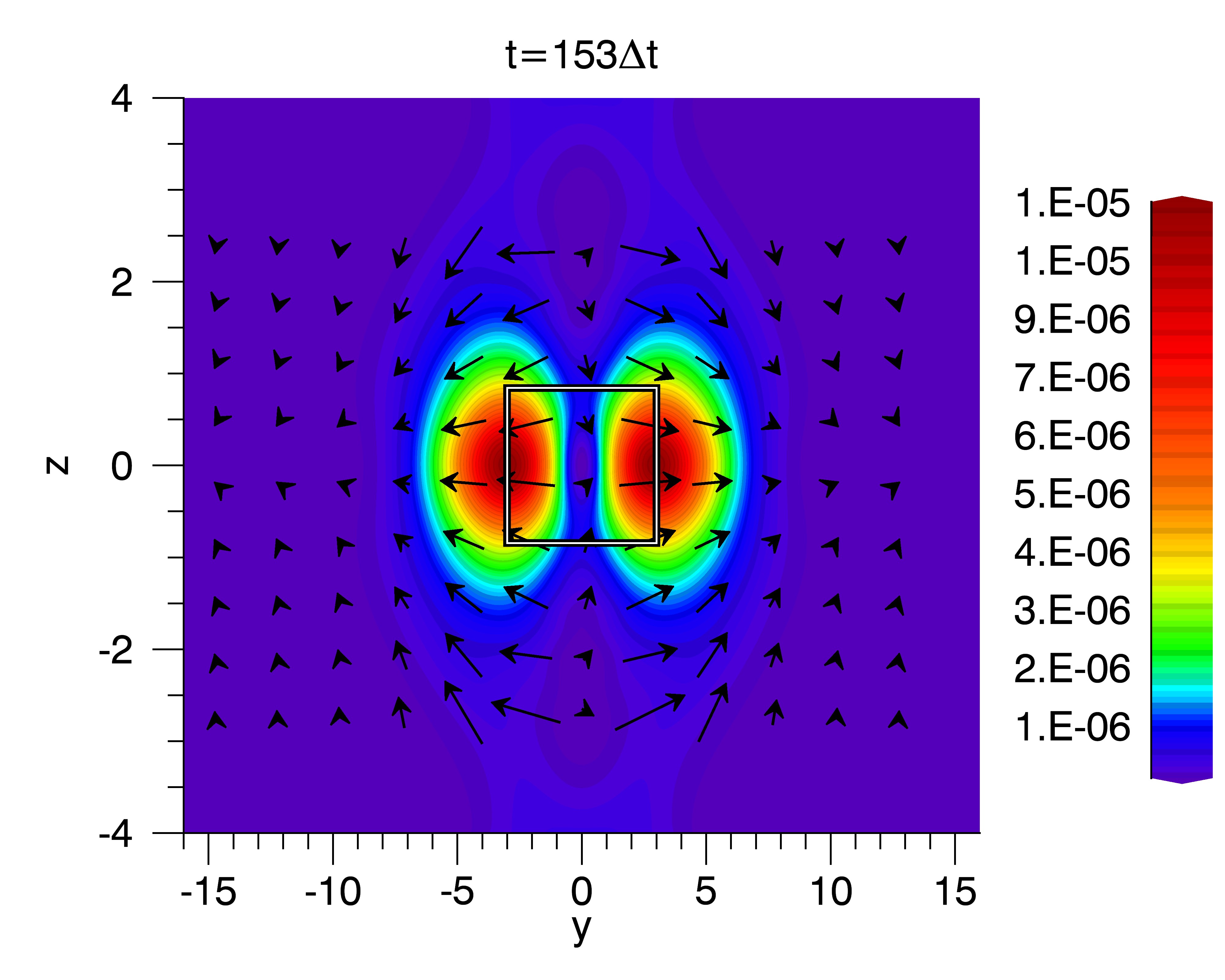}
\includegraphics[scale=0.04,clip=true, trim=2cm 2cm 1cm 3cm]{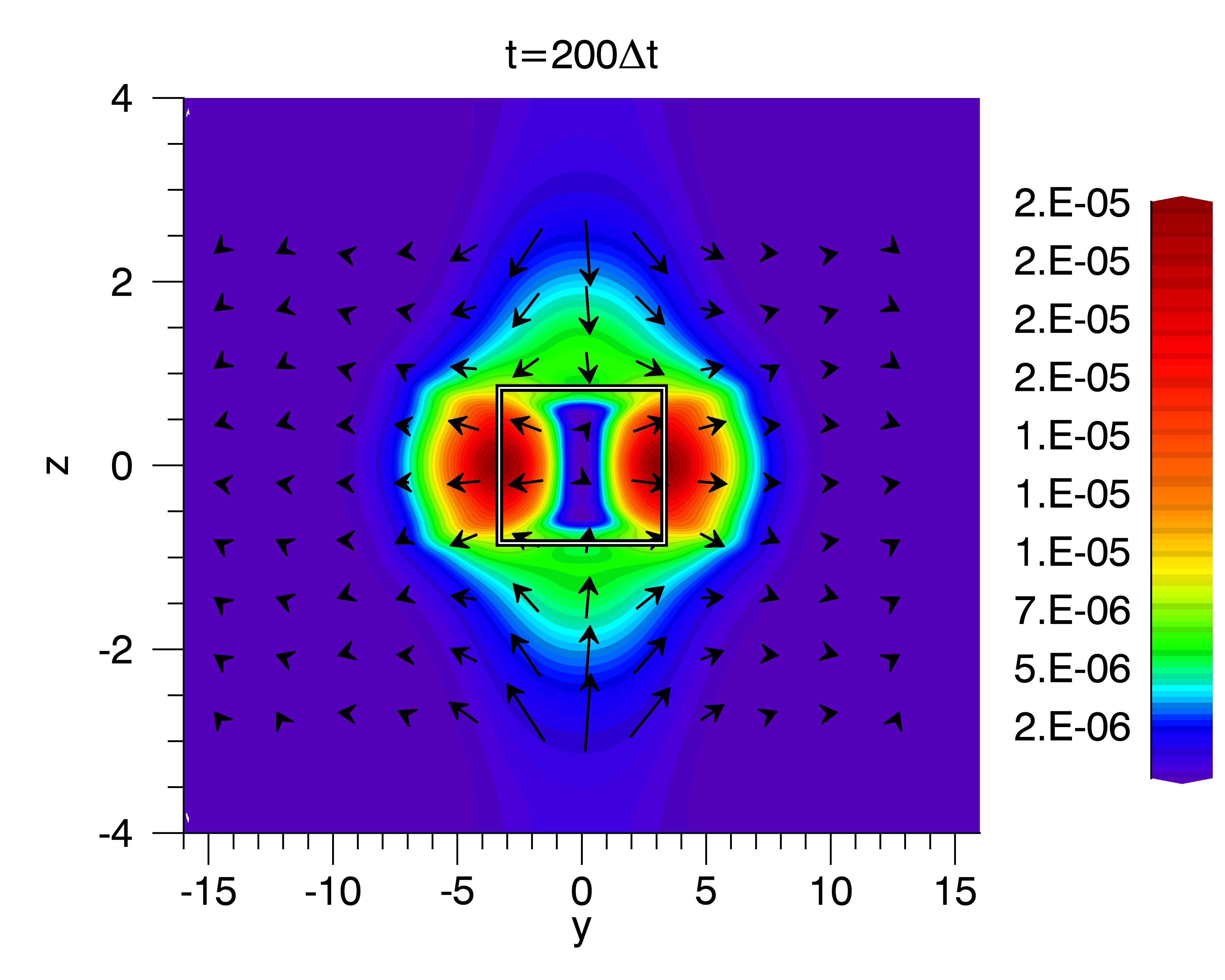}
\includegraphics[scale=0.04,clip=true, trim=2cm 2cm 1cm 3cm]{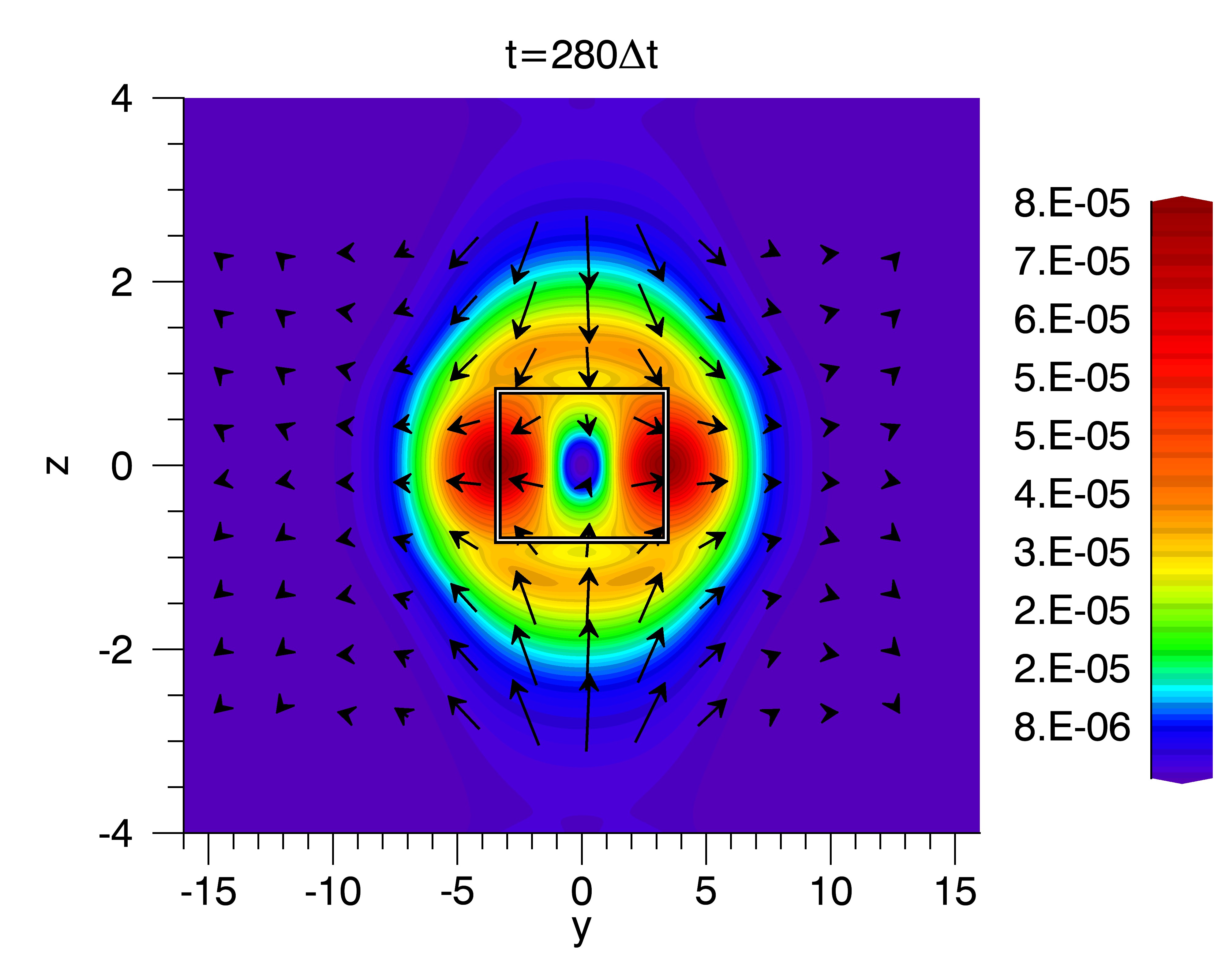}
\caption{Development phase shown in kinetic energy at times 153$\Delta$t, 200$\Delta$t and 280$\Delta$t for the coronal atmosphere. The white box frame denotes the reconnection region.}
\label{figfullydev}
\end{figure*}

\begin{figure*}
\centering
\includegraphics[scale=0.04,clip=true, trim=2cm 2cm 1cm 3cm]{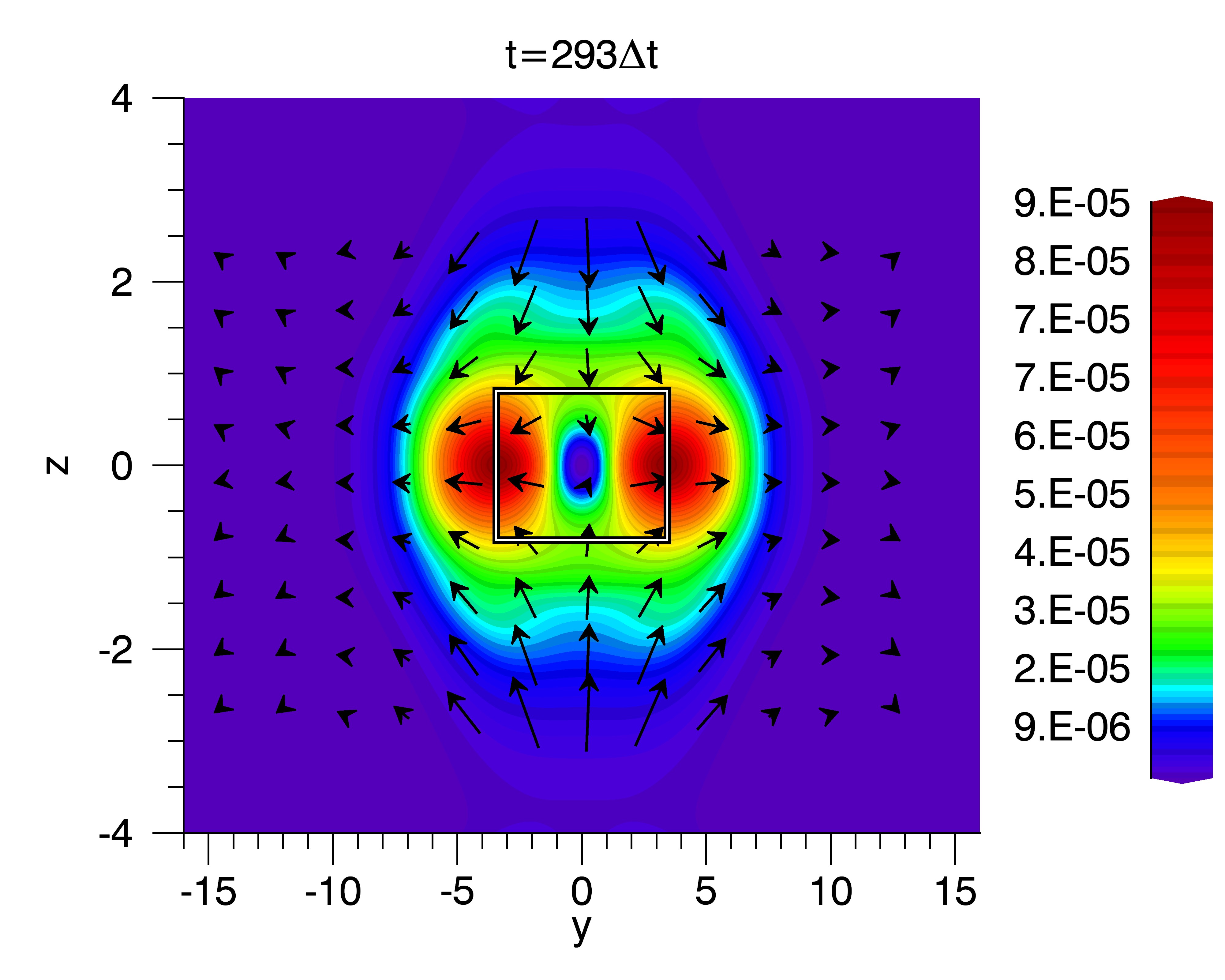}
\includegraphics[scale=0.04,clip=true, trim=2cm 2cm 1cm 3cm]{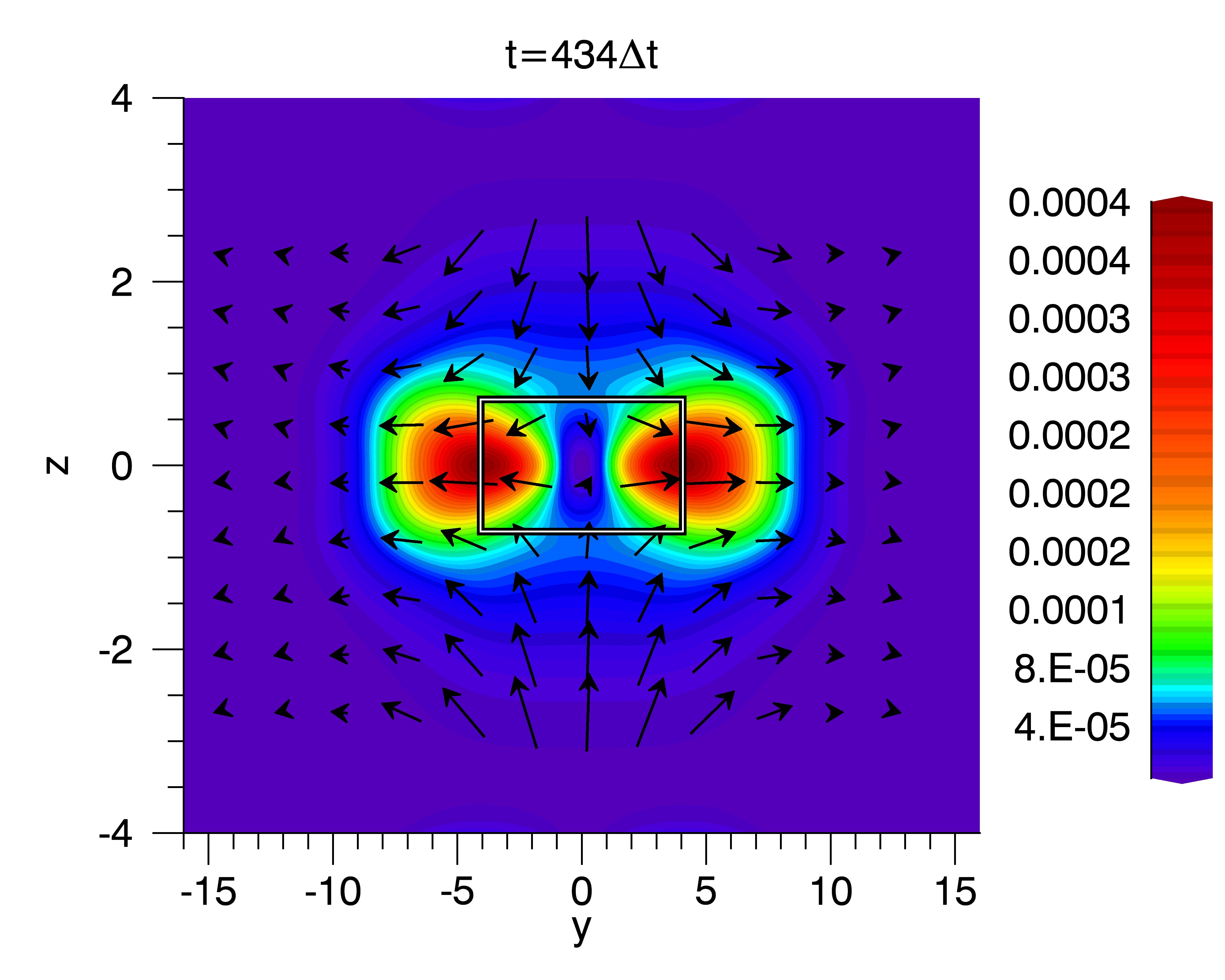}
\includegraphics[scale=0.04,clip=true, trim=2cm 2cm 1cm 3cm]{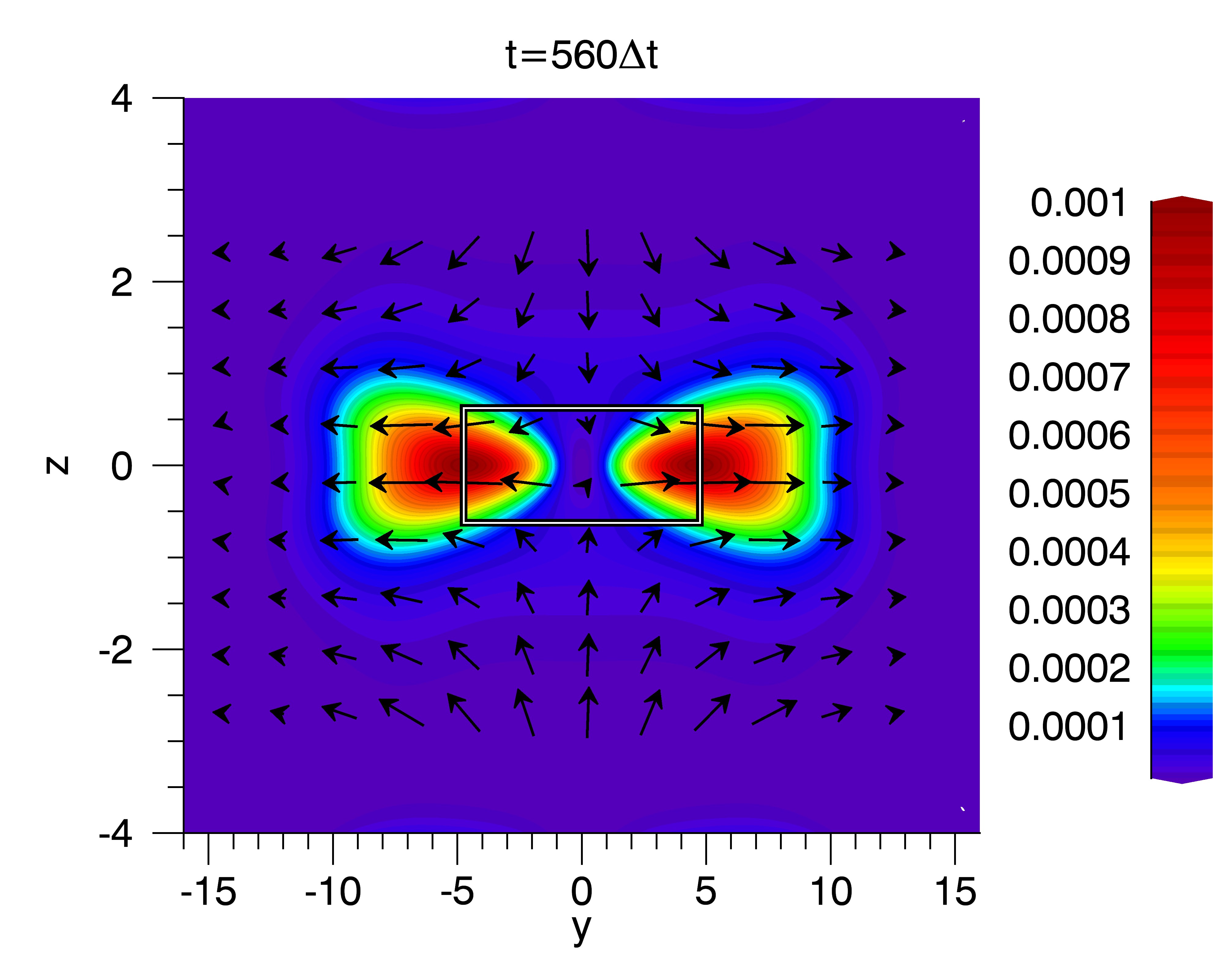}
\caption{Advection-dominated phase shown in kinetic energy at times 293$\Delta$t, 434$\Delta$t and 560$\Delta$t for the coronal atmosphere. The white box frame denotes the reconnection region.}
\label{figfullyrec}
\end{figure*}

The onset process can be separated into three phases: reflection (Figure \ref{figfullyref}), development (Figure \ref{figfullydev}) and advection-dominated  (Figure \ref{figfullyrec}). In this section each phase is discussed in order to understand the evolution of the system. More quantitative analysis of the coronal (reference) model is provided in subsequent sections.

\begin{figure*}
\centering
\includegraphics[scale=0.085,clip=true, trim=2cm 1cm 1cm 7cm]{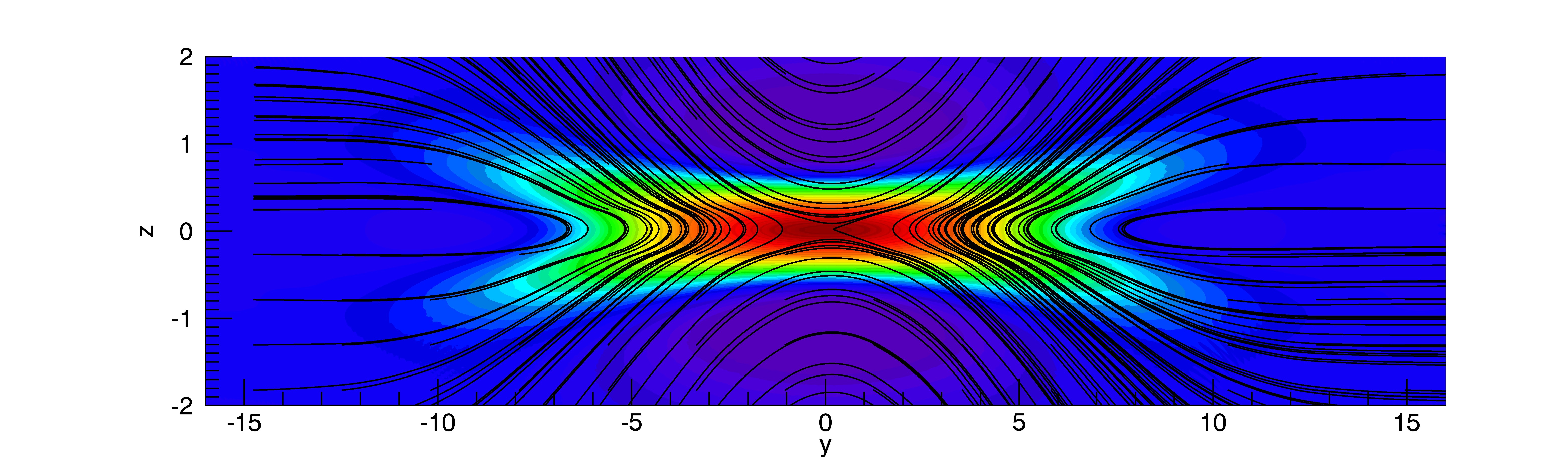}
\caption{Coloured contour of the perturbation current density $J_x -J_x (t=0)$ (red represents a large change, blue represents a small change) and streamlines of the normalised magnetic field at time $t=600\Delta t$.}
\label{figfullycur}
\end{figure*}

\subsection{Reflection phase}

The driven velocity propagates from $z=\pm 4$ at the local fast-mode speed towards the current sheet located at $z=0$.
The equilibrium Alfv\'en speed decreases towards the centre of the domain due to the increase in gas pressure and the decrease in magnetic field strength. The initial sound speed is constant across the domain because the temperature is constant. The fast-mode speed is a function of the equilibrium Alfv\'en speed and the sound speed and decreases towards the centre of the domain. Thus, the driven wave steepens as it approaches the centre of the Harris current sheet. The velocity amplitude of the driver is chosen to prevent this wave steepening into a shock. \citep[We note that Lare3D has shock-capturing capabilities, ][]{Arber2001}. 
When these wavefronts hit the centre of the Harris current sheet they trigger a pair of fast-mode reflections that propagate backwards towards the drivers (Figure \ref{figfullyref}). The majority of the kinetic energy of the fast-mode reflection is damped out in the damping region and has no significant impact on the dynamics in the current sheet. 

\subsection{Development phase}

In the wake of this fast-mode reflection, the current at the centre of the domain increases and the reconnection region begins to form (Figure \ref{figfullydev}). There is a small amount of outflow at the start of this phase, however there is no electric field increase associated with this. Therefore this is only a fluid process as a result of the driver, as opposed to a magnetic process. The current rises as a result of the inflow pushing magnetic field together. Towards the end of this stage, magnetic reconnection begins to accelerate the plasma and there is an associated rise in electric field. Note that there is uniform resistivity in the model so diffusion is always present. The magnetic field pile-up in the inflow region creates a larger current and allows plasma to be significantly accelerated by reconnection.

\subsection{Advection-dominated phase}

For late times, there is clear acceleration of the plasma exiting the reconnection region (Figure \ref{figfullyrec}). The reconnection reaches a linear phase in terms of reconnection region size, with $L$ increasing linearly with time and $\delta$ decreasing linearly with time. The reconnection event has qualities of both Sweet-Parker and Petschek models. The reconnection region is elongated and narrow like in Sweet-Parker reconnection, however there are weak slow-mode shocks that form on the interface between the inflow and outflow flow regions similar to Petschek. The inflow velocity is constant at late times, however the magnetic flux entering the reconnection region is increasing. This creates a reconnection process that behaves like the flux pile-up model of \cite{Priest1986}. This flux pile-up on the inflow region is also present in numerical simulations of the Taylor problem and acts to enhance the current sheet \citep[][]{Wang1996}. A colourmap of the perturbation current $J_x - J_x(t=0)$ and streamlines of magnetic field at time $t=600 \Delta t$ are shown in Figure \ref{figfullycur}. During this phase, the physics is dominated by the advection term in Ohm's law, as discussed in Section \ref{secvelex}.

\section{Velocity dependence}\label{sec_vel}

The coronal model in the previous section acts as our reference case and provides a qualitative analysis of the onset process. We now consider variation from this reference model to determine the role of various parameters on the onset of reconnection.
First the dependence on the magnitude of the velocity driver will be considered. Three different velocity magnitudes are presented here. The velocities have been chosen to prevent shocks occurring on the inflow. The driver Alfv\'en speed is shown is Table \ref{tabveltest}. The high driver considered here is the same as the reference coronal case in Section \ref{sec_fid}.

\begin{table}[!h]
\caption{Driver Alfv\'en Mach numbers for the velocity dependence test.}
     $$ 
         \begin{array}{p{0.5\linewidth}l}
            \hline
            \noalign{\smallskip}
            Driver name  &  \mbox{Alfv\'en Mach number}  \\
            \noalign{\smallskip}
            \hline
            \noalign{\smallskip}
            High & 0.003     \\
            Medium & 0.001             \\
            Low & 0.0003  \\
            \noalign{\smallskip}
            \hline
         \end{array}
     $$ 
\label{tabveltest}
\end{table}

\subsection{Evolution of the reconnection region}

\begin{figure*}[ht]
\begin{subfigure}{.45\textwidth}
(a)
\end{subfigure}
\begin{subfigure}{.45\textwidth}
(b)
\end{subfigure}
\centering
\begin{subfigure}{.48\textwidth}
\includegraphics[scale=0.55,clip=true, trim=3cm 7cm 2cm 8cm]{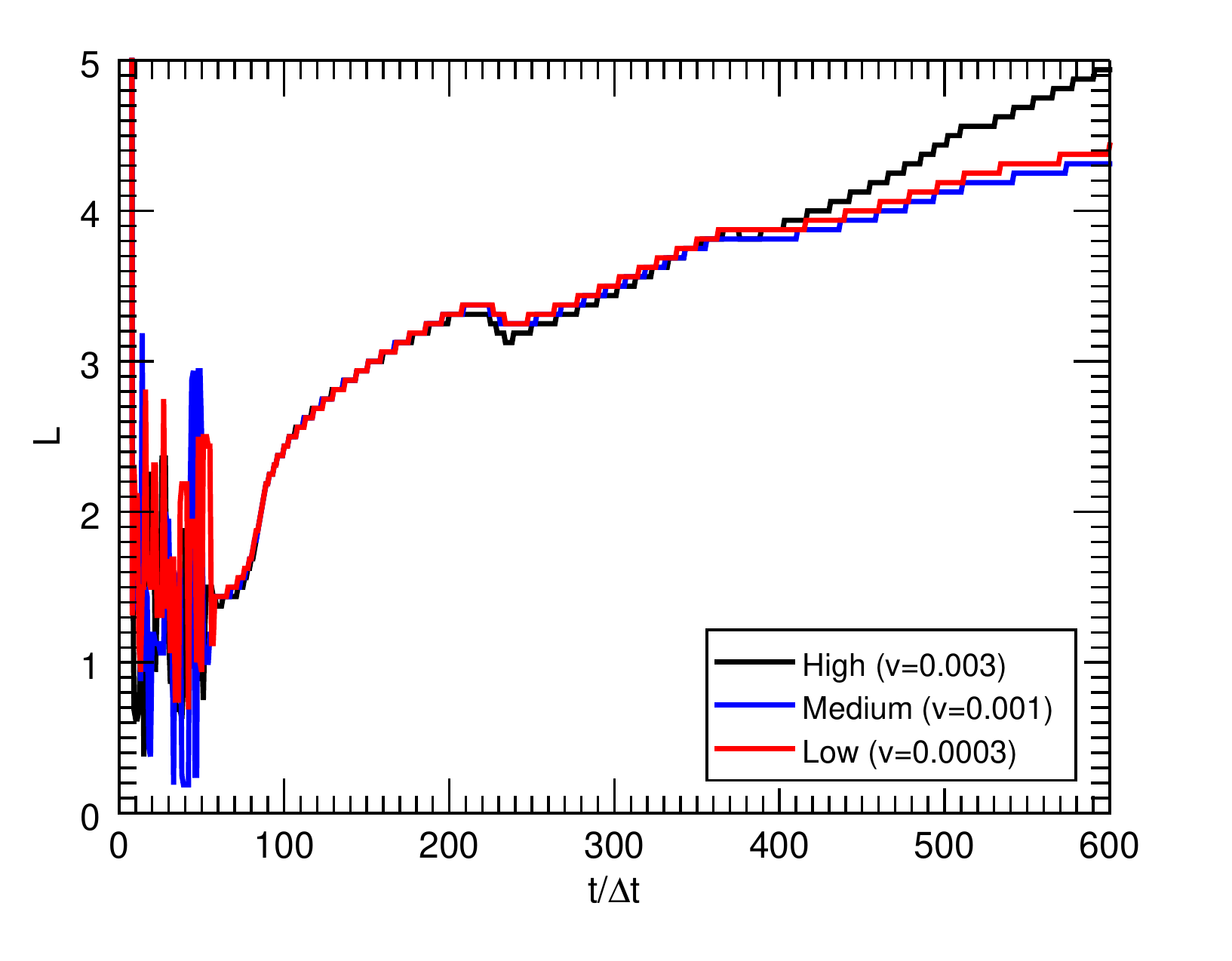}
\end{subfigure}
\begin{subfigure}{.48\textwidth}
\includegraphics[scale=0.55,clip=true, trim=3cm 7cm 2cm 8cm]{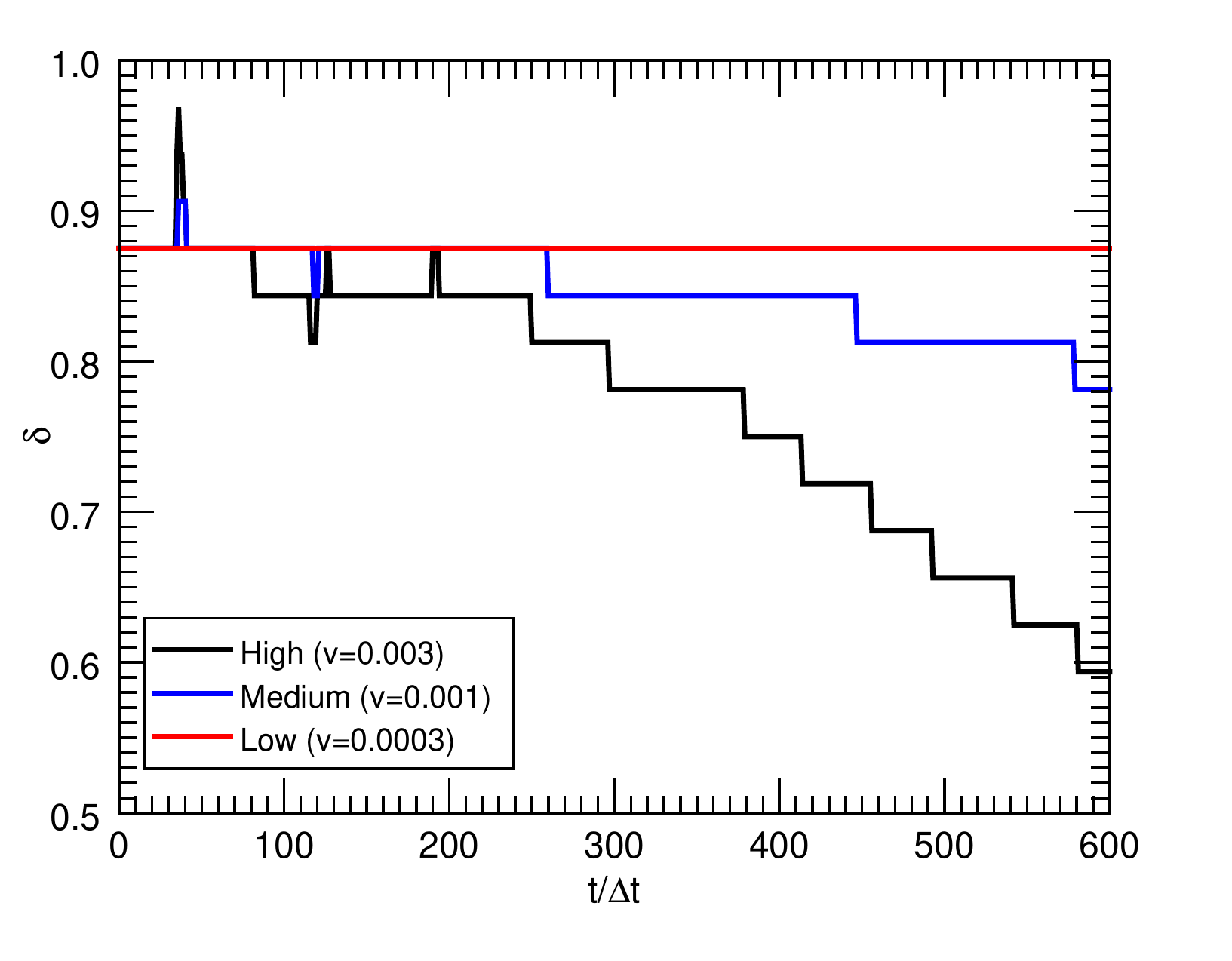}
\end{subfigure}
\caption{(a) Half-length $L$ and (b) half-width $\delta$ of the reconnection region for the coronal case. $L$ is calculated using the maximum outflow. $\delta$ is calculated as the HWHM of the current density. $L$ and $\delta$ are normalised by $l_0$ (Table \ref{tabbasepar1}).}
\label{figfullydif}
\end{figure*}

The half-width $\delta$ and half-length $L$ of the reconnection region are shown in Figure \ref{figfullydif}. After the initial development phase, the half-length $L$ of the reconnection region increases and the half-width $\delta$ decreases. All tested driver velocities influence $L$ similarly until approximately $t=400 \Delta t$. There are two linear post-development phases in the diffusion half-length $L$. The first between $250 \Delta t$ and $350 \Delta t$ during which all velocities yield the same growth rate for $L$. However after $t=400 \Delta t$ the high driver demonstrates a steeper growth rate for $L$. The diffusion half-width $\delta$ decreases linearly after $t=250 \Delta t$. The low driver velocity does not appear to affect $\delta$. 

The velocity driver acts to bring magnetic flux into the diffusion region, resulting in an increased current density in the inflow region. For the low velocity driver, the current density rises by only a small amount and hence the half-width $\delta$ (calculated as the HWHM of current density) remains constant. For the other velocity drivers, the current density increases sufficiently in the diffusion region to produce a smaller $\delta$.

\subsection{Electric field magnitude and evolution}\label{secvelex}

The maximum electric field occurs at the centre of the domain when diffusion dominates. When advection dominates, the maximum electric field moves along the line $z=0$ with the outflow. Note however the electric field variation along this line is small inside the diffusion region (significantly less than an order of magnitude).

\begin{figure}[ht]
\centering
\includegraphics[scale=0.55,clip=true, trim=2cm 7cm 2cm 8cm]{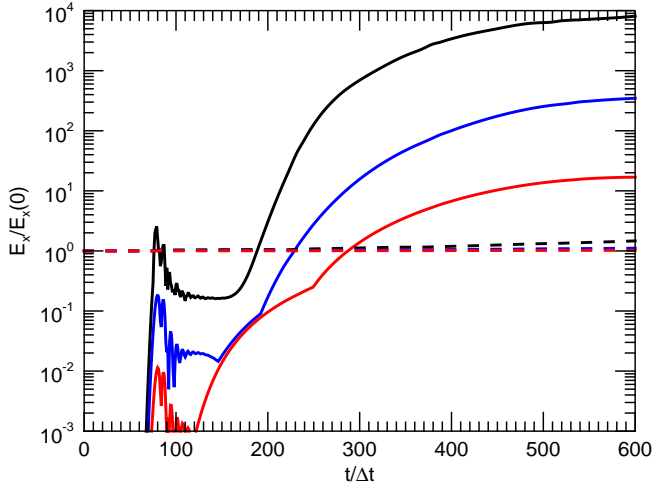}
\caption{Components of the Ohm's law: $| \textbf{v} \times \textbf{B} |$ (solid line) and $\eta J_x$ (dashed line), normalised by the total electric field at time $t=0$. The electric field value is chosen as the maximum value inside the diffusion region. The colour code of the lines is the same as in Figure \ref{figfullydif}.}
\label{figfullyex}
\end{figure}

The maximum electric field in the reconnection region shows a significant difference between the three different velocity drivers in terms of magnitudes. However, all have reasonably similar behaviour during the reflection and reconnection phases, see Figure \ref{figfullyex}. There is a sudden change in magnitude of the electric field that is dependent on the magnitude of the velocity driver. The electric field increases by approximately four, two and one orders magnitude for the high, medium and low drivers respectively. By looking at the components of the electric field at the location of the maximum electric field in Figure \ref{figfullyex}, one can see that the steep gradient change occurs in the $\textbf{v} \times \textbf{B}$ term. This implies the advection process dominates at late times. $\eta$ is small in this case so the diffusion term is also small. The $\textbf{v} \times \textbf{B}$ term shows the three distinct phases: reflection phase, development phase and advection-dominated phase. 

During the reflection phase, the wavefront from the driver hits the centreline (at time $t=80 \Delta t$) and causes a fast-mode reflection. In the wake of this reflection the current sheet fluctuates in width towards a new equilibrium. This same process occurs for all three velocities tested. These fluctuations are present in the $\textbf{v} \times \textbf{B}$ component of Ohm's law, shown in Figure \ref{figfullyex} by the damped oscillations between $t=80 \Delta t$ and $100 \Delta t$. 

One can also identify the end of the development phase in Figure \ref{figfullyex} as the point where the $\textbf{v} \times \textbf{B}$ term becomes larger than the $\eta \textbf{J}$ term. This occurs at a different time in all cases. Here some magnetic reconnection is occurring, however the process appears to be different in each case. To understand this, one must consider the components of $\textbf{v} \times \textbf{B}$ in the $x$-direction, that is $v_y B_z$ and $v_z B_y$. The magnetic field in the $y-$direction is far stronger than the $z-$direction and hence the effects of $v_z$ are magnified when looking at $\textbf{v} \times \textbf{B}$. The high driver shows a plateau in the development region, at approximately times $t = 120 \Delta t$ to $200 \Delta t$ in Figure \ref{figfullyex}. The pressure of the incoming velocity is sufficient to result in a velocity directed predominantly in the $y-$direction at the location of the maximum electric field. The low velocity driver increases in $\textbf{v} \times \textbf{B}$ during the development phase. This is because the pressure from the incoming velocity is insufficient to direct the flow in the $y$-direction. Instead the velocity flows predominantly along the magnetic fieldlines. The medium driver is a combination of these two effects. 

During the advection-dominated phase, the $\textbf{v} \times \textbf{B}$ term dominates the diffusion term and slow-mode shocks begin to form. The difference between the magnitudes of the advection and diffusion terms is dependent on the driver velocity. For the high driver the advection term is approximately four orders larger than the diffusion term. A larger inflow velocity produces a larger reconnection signature since more magnetic flux is transported into the reconnection region.

\subsection{Resistivity dependence}\label{sec_res}

A diffusion test was performed on the simulation grid determining that the numerical resistivity is of order $\eta _{n} \approx 10^{-8}$ $\Omega$m (see Appendix \ref{appa}), that is far smaller than the Spitzer resistivity for the coronal case $\eta = 10^{-6}$ $\Omega$m (see Table \ref{tabbasepar1}). However, performing the simulation with a Spitzer resistivity of $\eta =10^{-3}$ $\Omega$m and $\eta =10^{-4.5}$ $\Omega$m did not significantly change the structure of the current density. The reconnected magnetic field along the line $z=0$ does increase for the higher resistivity however remains several orders of magnitude smaller than the inflow magnetic field. This shows that changing the resistivity does have an effect on the amount of reconnected magnetic field, however the reconnected flux remains too small to have an effect on other parameters. The velocity driver carries magnetic field towards the diffusion region and hence, the rate at which magnetic flux enters the diffusion region is determined by the velocity amplitude. This acts as a limiting factor on the reconnection rate during the onset of magnetic reconnection. Towards the end of the simulation, the advection term stabilises but the current term is still increasing. One would expect that the diffusion term $\eta J$ would eventually dominate the physics of the reconnection process as time increases beyond our simulation. It was not possible to extend the simulation time due to numerical stability issues. A weak dependence of the resistivity is also present in results from the Taylor problem \citep[e.g.][]{Wang1996}.

\section{Plasma-$\beta$ dependence} \label{sec_beta}
To investigate the effect of the plasma-$\beta$, the high driver (Alfv\'en Mach number $0.003$) from the previous section is used (note that this is also used in the reference coronal model in Section \ref{sec_fid}). Three plasma-$\beta$ values have been tested: $0.002, 0.1$ and $1$. The change in plasma-$\beta$ changes the density distribution (see Eq. (\ref{den_dist})) and hence the propagation time of the driven velocity. This changes the time at which the wavefronts collide. Note that the change in plasma-$\beta$ affects the initial equilibrium sound speed, and thus the fast-mode propagation speed, via the density and wave propagation time.

\subsection{Reconnection region}

\begin{figure*}[ht]
\begin{subfigure}{.45\textwidth}
(a)
\end{subfigure}
\begin{subfigure}{.45\textwidth}
(b)
\end{subfigure}
\centering
\begin{subfigure}{.48\textwidth}
\includegraphics[scale=0.45,clip=true, trim=1cm 6cm 1cm 7cm]{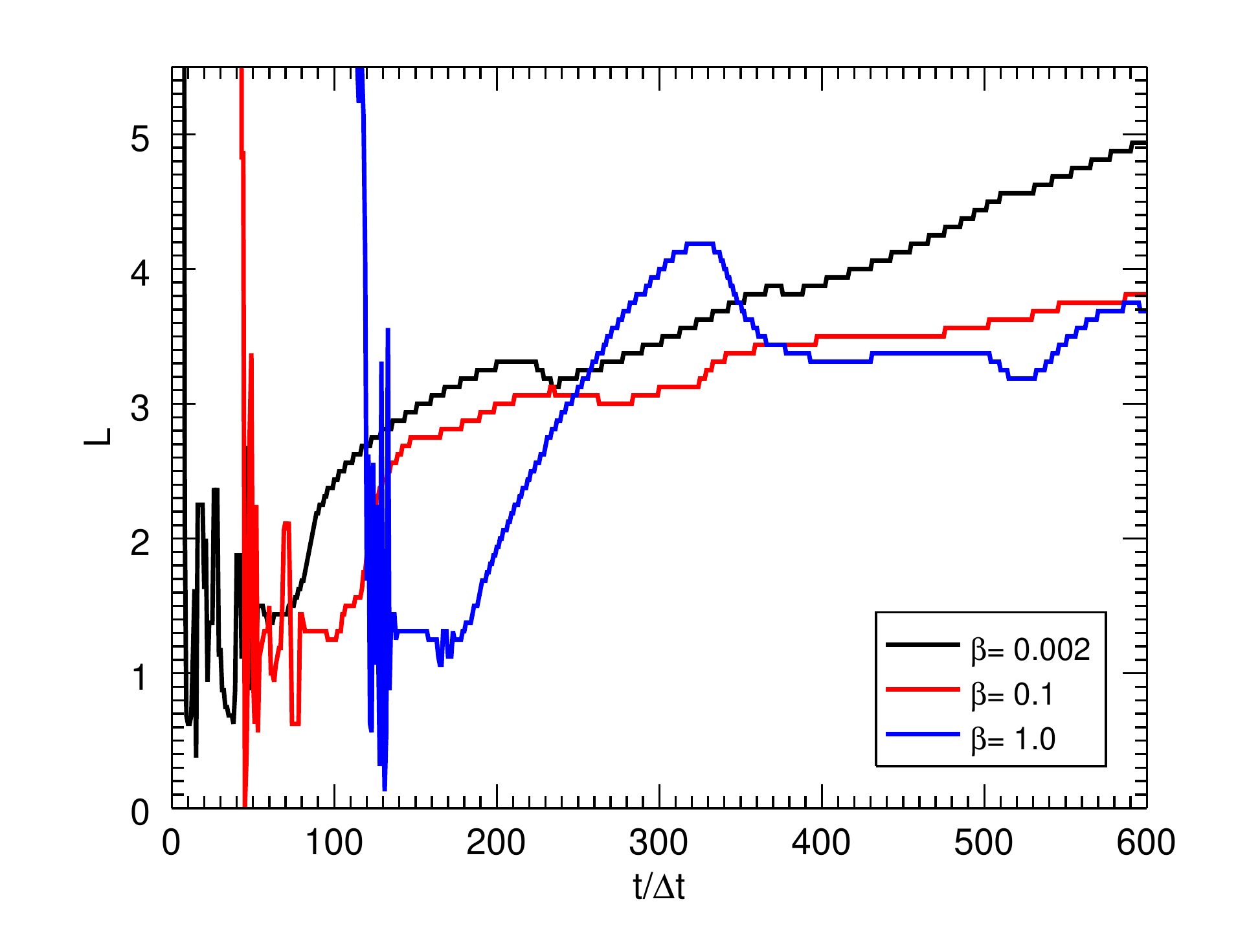}
\end{subfigure}
\begin{subfigure}{.48\textwidth}
\includegraphics[scale=0.45,clip=true, trim=1cm 6cm 1cm 7cm]{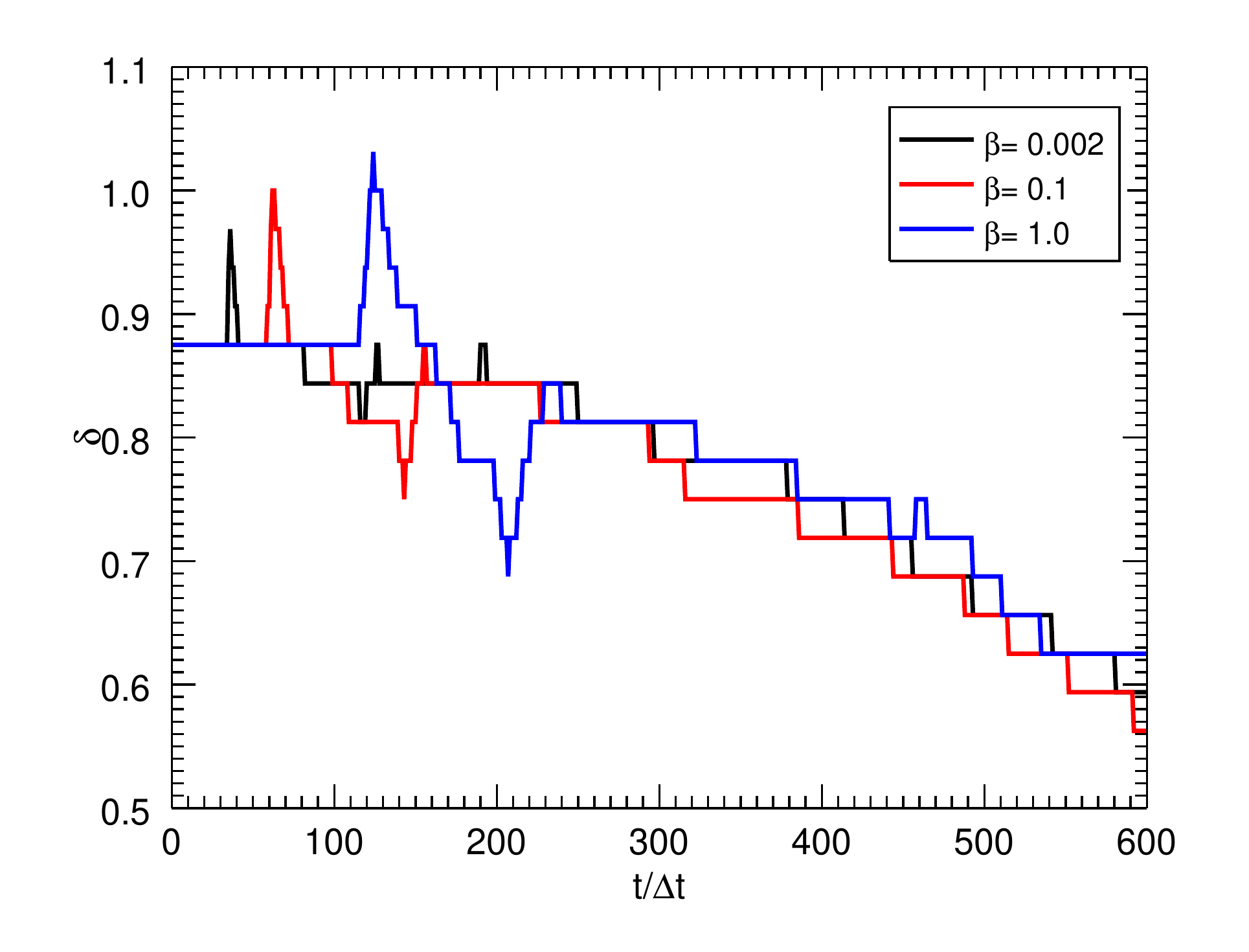}
\end{subfigure}
\caption{(a) Half-length $L$ and (b) half-width $\delta$ for different plasma-$\beta$ values.}
\label{figbetadelta}
\end{figure*}

As before, a reconnection region of width $2\delta$ and length $2L$ can be fitted to the data. The half-width $\delta$ and half-length $L$ are shown in Figure \ref{figbetadelta}.

The half-width $\delta$ shows nearly identical values for late times. The main differences appear to be the magnitude of $\delta$ during the reflection phase, when the wavefront hits the current sheet. However all three cases show the same qualitative behaviour. The absolute magnitude of the velocity driver is the same across the three cases and hence the rate at which magnetic flux is entering the diffusion region is the same. This results in similar increases in current density in the inflow region and hence similar behaviour in $\delta$.

The half-length $L$ shows differences between the three tested plasma-$\beta$ values. All three cases show a general increasing trend however the rates of this increase vary. In the case where plasma-$\beta = 0.002$, $L$ increases in an approximately linear fashion. For the high plasma-$\beta$ case, $\beta =1$, the half-length $L$ is relatively constant after the initial spike due to the waves colliding. $L$ starts to increase for this case towards the end of the simulation. As the plasma-$\beta$ increases, plasma pressure becomes more important and the outflow velocity is determined more by fluid motion, resulting in a slower increase in $L$ for higher plasma-$\beta$ values.

\subsection{Electric field: variation with plasma-$\beta$}

\begin{figure}[!h]
\centering
\includegraphics[scale=0.45,clip=true, trim=1cm 6cm 2cm 7cm]{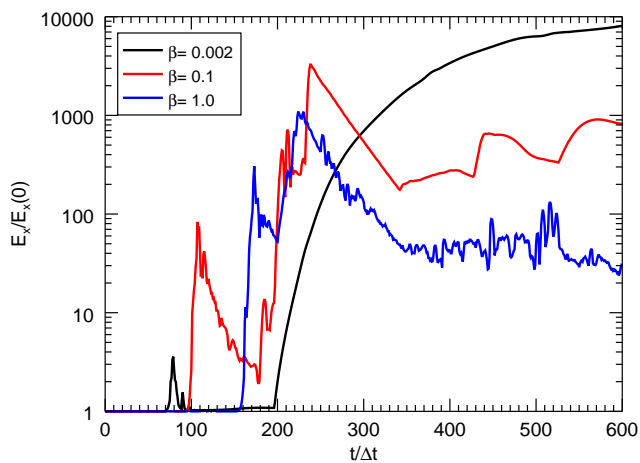}
\caption{Maximum normalised electric field in the reconnection region for different plasma-$\beta$ values.}
\label{figbetaex}
\end{figure}

The maximum electric field in the reconnection region is shown in Figure \ref{figbetaex}. There are significant differences between the electric field signature for different plasma-$\beta$ values.

The low plasma-$\beta$ case ($\beta = 0.002$) was analysed in Section \ref{sec_fid} as the reference coronal model and Section \ref{sec_vel} as the high velocity case. A peak in electric field occurs during the reflection phase, when the two wavefronts collide at approximately $t=80 \Delta t$. After this, the electric field is relatively constant throughout the development phase. Finally there is a rise of four orders of magnitude in electric field as the reconnection phase starts up at approximately time $t= 200 \Delta t$ in Figure \ref{figbetaex}. This is due to the $\textbf{v} \times \textbf{B}$ term in Ohm's law becoming far larger than the $\eta \textbf{J}$ term as was seen in Figure \ref{figfullyex} and discussed in Section \ref{secvelex}.

The other plasma-$\beta$ values behave quite differently. As the plasma-$\beta$ value increases, the plasma pressure becomes more important. This results in a larger electric field spike when the two wavefronts collide. Following this, the behaviour is different than in the reference model (black line in Figure \ref{figbetaex}). The development phase occurs at a later time and has multiple peaks as oppose to being fairly constant. The advection-dominated phase is also very different from the reference coronal model (Section \ref{sec_fid}); there is no large, smooth increase, instead the electric field is fairly constant and noisy. For the plasma-$\beta =1$ case the advection term is far larger than the diffusion term. For the plasma-$\beta =0.1$ case, the advection and diffusion terms are of similar orders of magnitude.

Increasing the plasma-$\beta$ appears to inhibit the reconnection. The electric field only demonstrates a sharp rise, indicative of reconnection, for high plasma-$\beta$ values when the inflow velocity approaches the Alfv\'en speed.  

\section{Ambipolar diffusion} \label{sec_ambi}

In the solar chromosphere, the plasma is partially ionised. This introduces Cowling resistivity $\eta _{\perp}$ that acts perpendicular to the magnetic field and is calculated by Equation (\ref{eqneta}). $\eta _\perp$ appears in the internal energy equation (Equation \ref{eqnieeqn}) and Ohm's law (Equation \ref{eqnohmslaw}). Ambipolar diffusion refers to the $\eta _\perp J_\perp$ term in Ohm's law (Equation \ref{eqnohmslaw}).

To investigate the effects of Cowling resistivity, simulations were performed using chromospheric conditions (see Table \ref{tabbasepar2}). Three initial temperature values have been chosen at 7200, 9200 and 10300 K such that the respective neutral fraction $\xi _n$ is 0.9, 0.5 and 0.1. The driver velocity is chosen to avoid shocks in the inflow region at all three temperatures. We note that the change in temperature changes the propagation speed and hence the time at which the wavefronts collide.

\begin{table}[!h]
\caption{Chromospheric parameters}
     $$ 
         \begin{array}{p{0.5\linewidth}l}
            \hline
            \noalign{\smallskip}
            Parameter  [Units]    &  \mbox{Value} \\
            \noalign{\smallskip}
            \hline
            \noalign{\smallskip}
            Plasma-$\beta$ & 0.1     \\
            Temperature [K]& 7200, 9200 \mbox{ and } 10300             \\
            Corresponding $\xi _n$ & 0.9, 0.5 \mbox{ and } 0.1 \\
            Resistivity [$\Omega$m]& 10^{-3}  \\
            \noalign{\smallskip}
            \hline
         \end{array}
     $$ 
\label{tabbasepar2}
\end{table}

\subsection{Reconnection region}
\begin{figure*}[!ht]
\begin{subfigure}{.45\textwidth}
(a)
\end{subfigure}
\begin{subfigure}{.45\textwidth}
(b)
\end{subfigure}\\
\centering
\begin{subfigure}{.48\textwidth}
\includegraphics[scale=0.55,clip=true, trim=3cm 7cm 2cm 8cm]{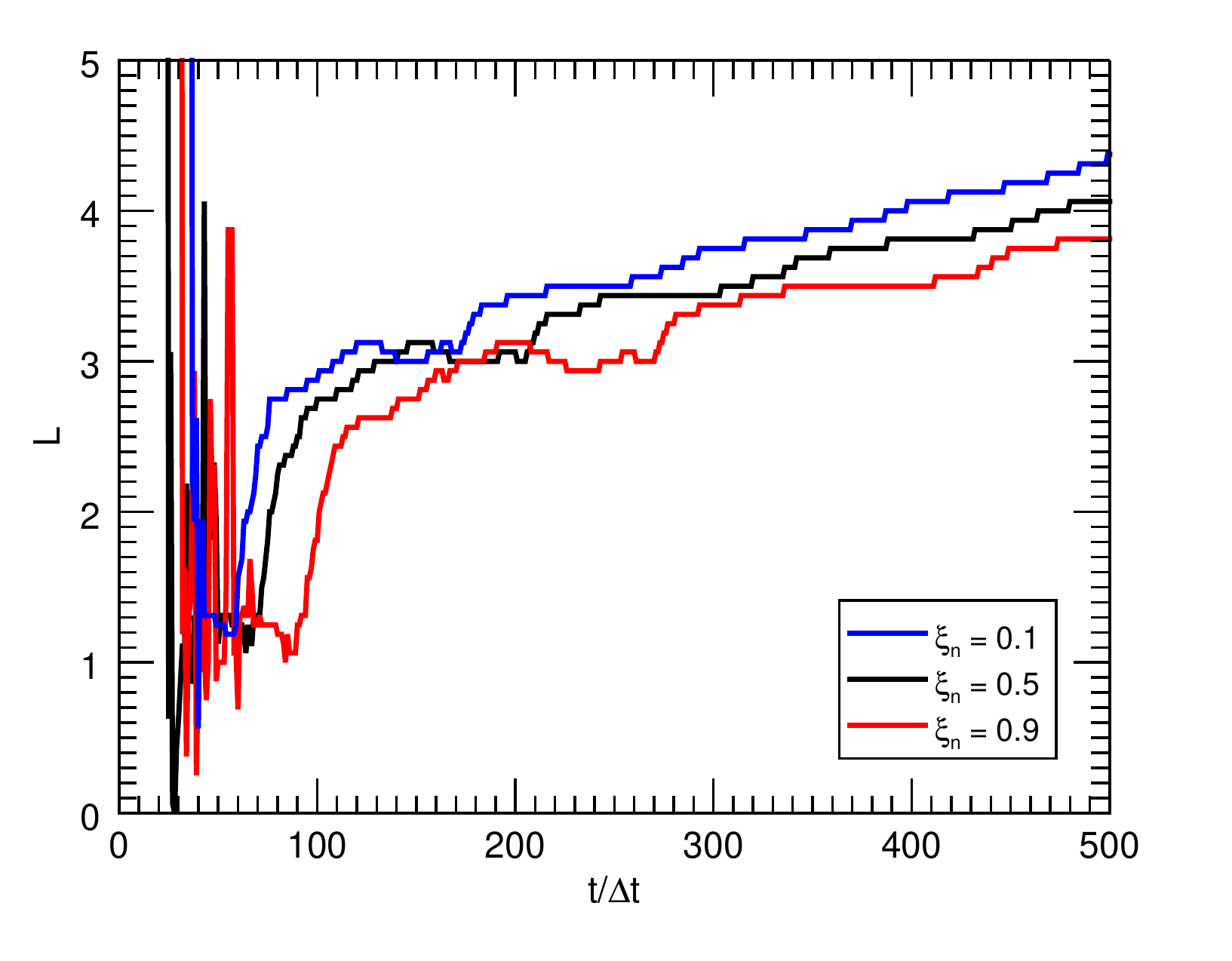}
\end{subfigure}
\begin{subfigure}{.48\textwidth}
\includegraphics[scale=0.55,clip=true, trim=3cm 7cm 2cm 8cm]{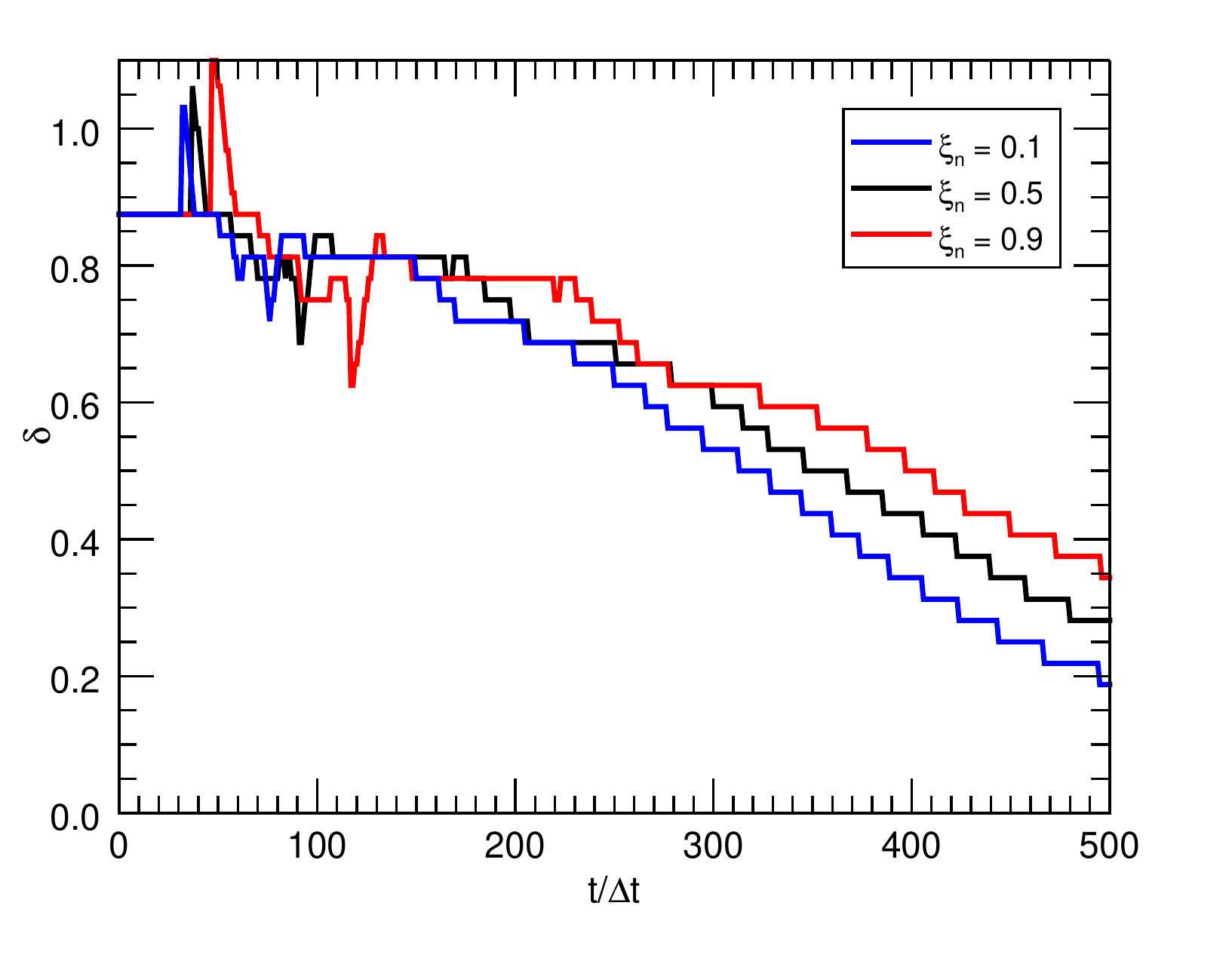}
\end{subfigure}
\caption{(a) Half-length $L$ and (b) half-width $\delta$ of the reconnection region with three different neutral fractions $\xi _n$. $L$ is calculated using the maximum outflow. $\delta$ is calculated as the HWHM of $J_x$.}
\label{figambidif}
\end{figure*}

The development and evolution of the reconnection region (half-length $L$ and half-width $\delta$) are shown in Figure \ref{figambidif}. At late times, all three cases show a linear increase in $L$ and a linear decrease in $\delta$. There is only minimal difference between the considered cases in terms of the reconnection region half-width and half-length. The atmospheric conditions are fairly similar in the three cases so the behaviour of the start-up of reconnection is similar across the three cases, that is a narrowing and elongating reconnection region with comparable gradients across the three cases.

Previous results indicate a correlation between the current layer thickness and neutral fraction \citep[e.g.][]{Yamada2006}. Our results in Figure \ref{figambidif}(b) show similar rates of change in the width $\delta$ during the onset of reconnection. This difference in these results originates from the stage of reconnection considered. Ambipolar diffusion is usually considered in the context of a tearing mode instability, where reconnection is achieved directly using an interior trigger \citep[e.g.][]{Zweibel1989}. In this paper we consider the early onset of reconnection, using an exterior trigger in the form of a velocity driver. It was established in \S \ref{sec_res} that the resistivity plays a minimal role in the onset of reconnection in our configuration and since the ambipolar diffusion manifests in the equations as an additional resistivity term one would expect that the ionisation fraction would have little role in the onset of reconnection. As time advances, resistivity begins to play a more dominant role and the effects of partial ionisation would become more pronounced. Note that outside the reconnection region, ambipolar diffusion shapes the inflowing current density (\S \ref{sec:currdist}).

\subsection{Electric field}
\begin{figure}[ht]
\centering
\includegraphics[scale=0.55,clip=true, trim=2cm 7cm 2cm 8cm]{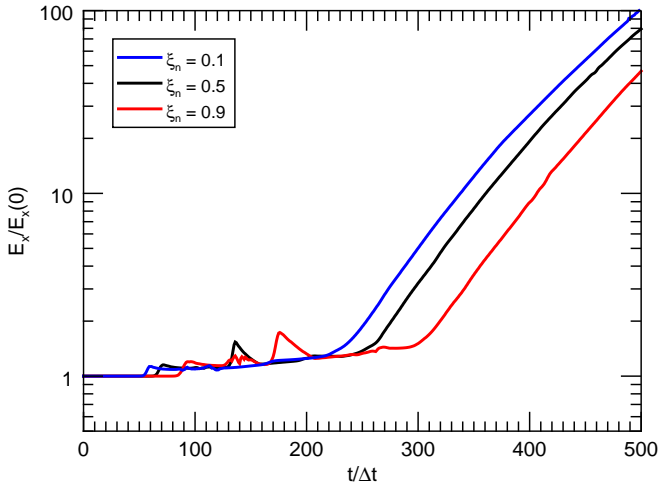}
\caption{Normalised electric field for different neutral fractions $\xi _n$.}
\label{figambiex}
\end{figure}

Figure \ref{figambiex} shows the evolution of the  maximum electric field in the reconnection region. Again there is very little difference between the cases. All three cases show near identical behaviour, both quantitatively and qualitatively. The time off-set of the results is due to the difference in temperature altering the propagation speed of the driven wavefronts, and thus the time at which the wavefronts collide. The large increase in electric field occurs when the $\textbf{v} \times \textbf{B}$ term becomes larger than the $\eta \textbf{J}$ term in Ohm's law. 

\subsection{Current distribution} \label{sec:currdist}
\begin{figure}[ht]
\centering
\includegraphics[scale=0.55,clip=true, trim=2cm 7cm 2cm 8cm]{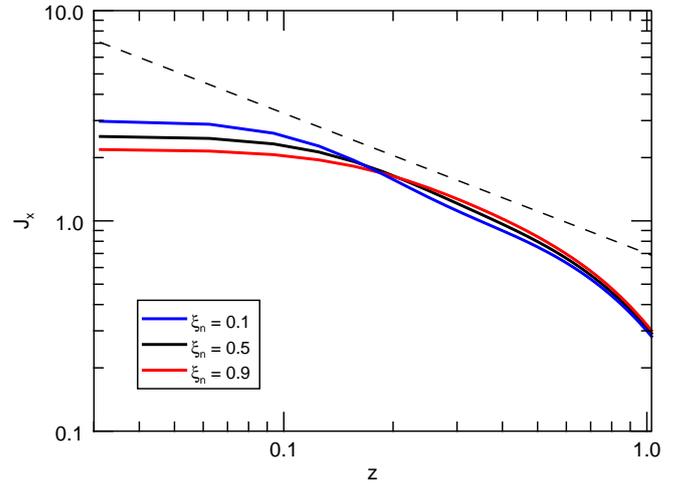}
\caption{Current $J_x$ for different neutral fractions $\xi _n$. The dashed line represents the $-2/3$ scaling law of \cite{Brandenburg1994}.}
\label{figambisca}
\end{figure}

For a partially-ionised plasma, the distribution of the current should follow a power law of $-2/3$ \citep{Brandenburg1994}. This power law region allows sharper features to form in partially-ionised plasmas, for example the solar chromosphere. Figure \ref{figambisca} shows the distribution of the current $J_x$ in the $z$-direction at time $t=500 \Delta t$ along the line $y=0$, that is the distribution of the current along the inflow direction. At this time the diffusion region is located at $z \approx 0.2,0.3\mbox{ and }0.4$ for $\xi _n = 0.1,0.5 \mbox{ and }0.9$ respectively (Figure \ref{figambidif}(b)). All three cases demonstrate a power law region with roughly the expected $-2/3$ gradient. This indicates that the inclusion of partial ionisation alters the distribution of the current along the inflow direction. In contrast, there is no power-law region in the distribution of the current for the coronal case. Figure \ref{figambisca} also shows that there are differences between the distribution of current for different neutral fractions. As time advances, one would expect these to become more pronounced as the resistivity begins to play a larger role in the reconnection.

\begin{figure*}[!ht]
\begin{subfigure}{.45\textwidth}
(a)
\end{subfigure}
\begin{subfigure}{.45\textwidth}
(b)
\end{subfigure}\\
\centering
\begin{subfigure}{.48\textwidth}
\centering
\includegraphics[scale=0.55,clip=true, trim=3cm 7cm 2cm 8cm]{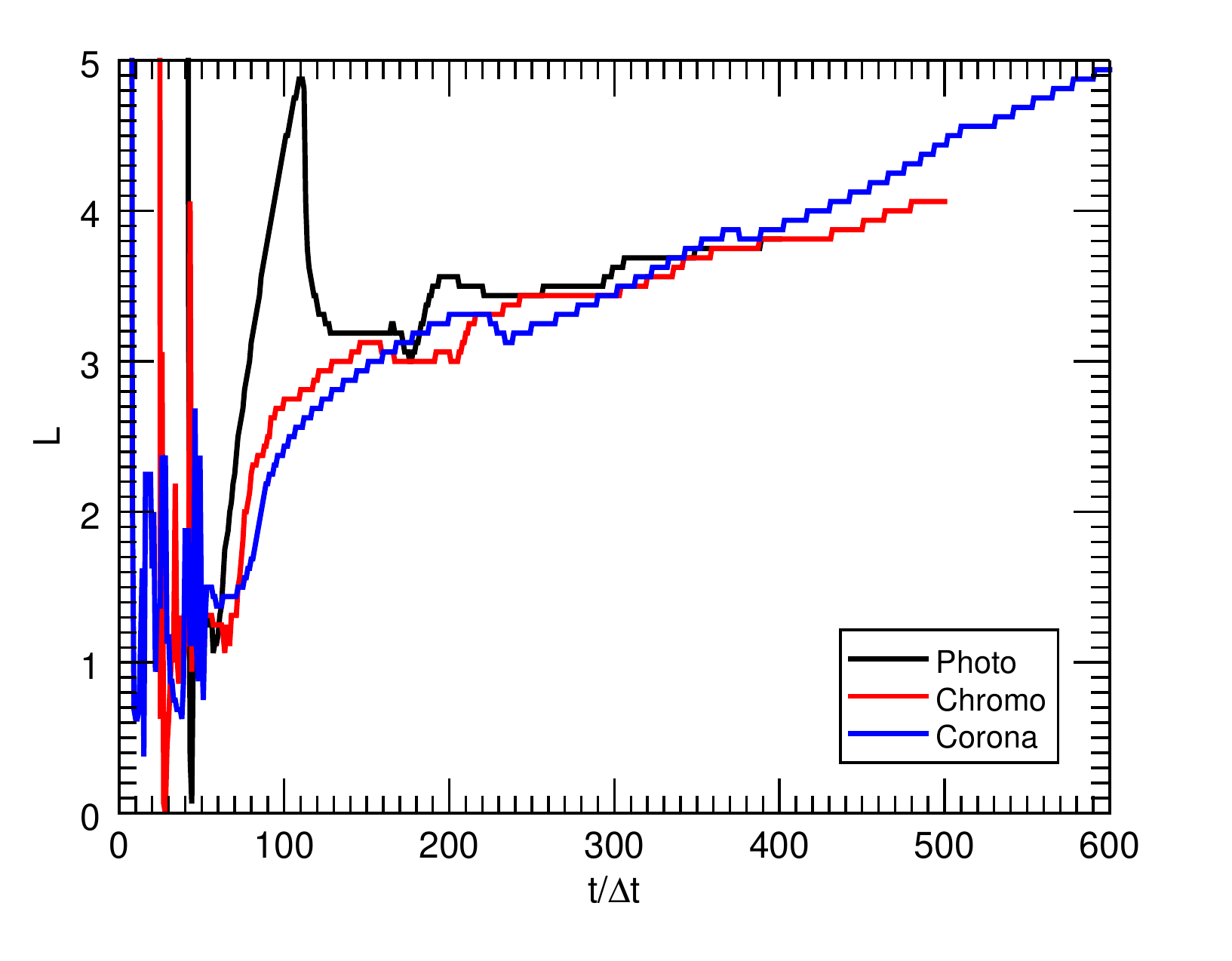}
\end{subfigure}
\begin{subfigure}{.48\textwidth}
\centering
\includegraphics[scale=0.55,clip=true, trim=3cm 7cm 2cm 8cm]{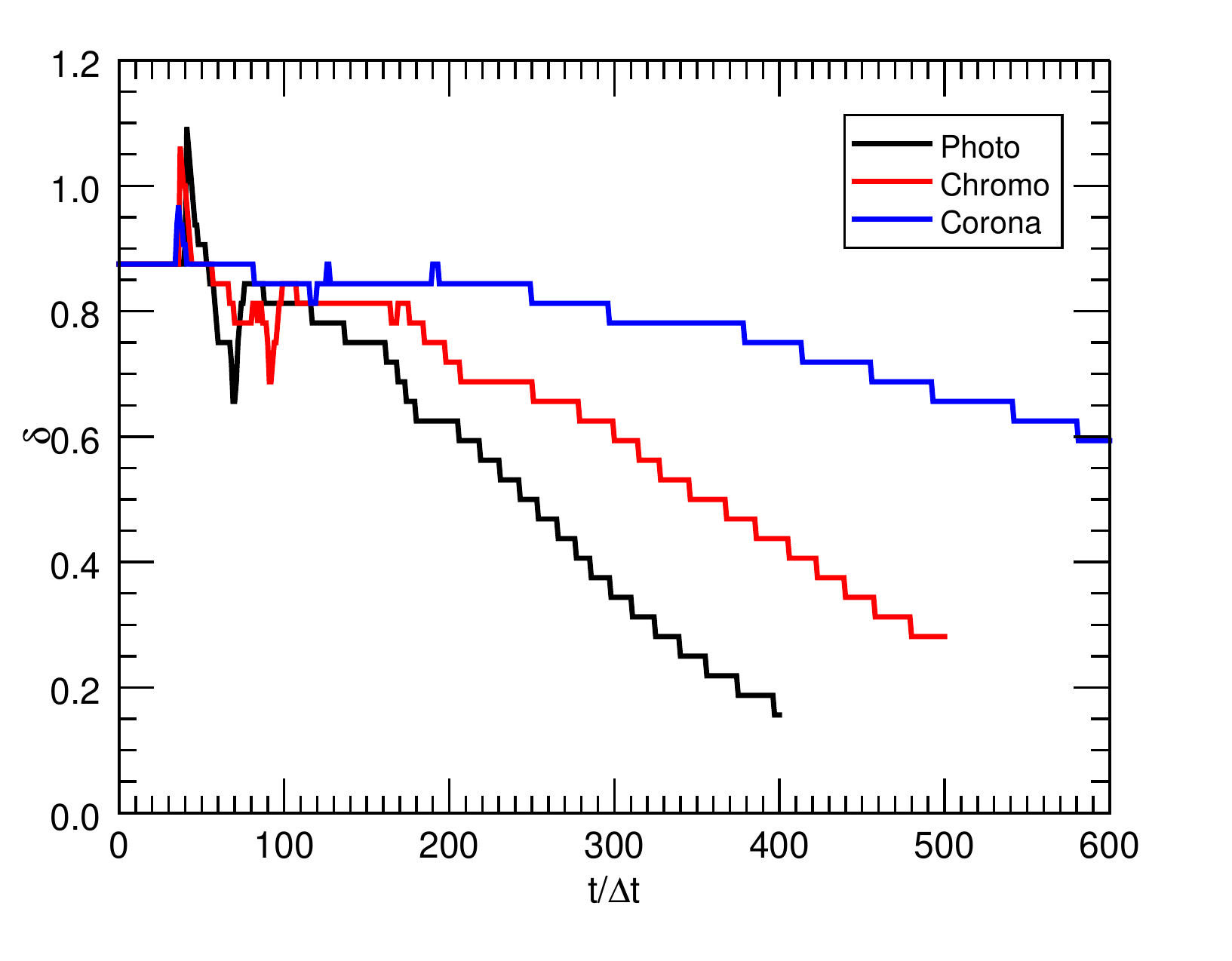}\end{subfigure}
\caption{(a) Half-length $L$ and (b) half-width $\delta$ of the reconnection region for the three atmospheric heights.}
\label{figcomdif}
\end{figure*}

\begin{figure}[!h]
\centering
\includegraphics[scale=0.55,clip=true, trim=2cm 7cm 2cm 8cm]{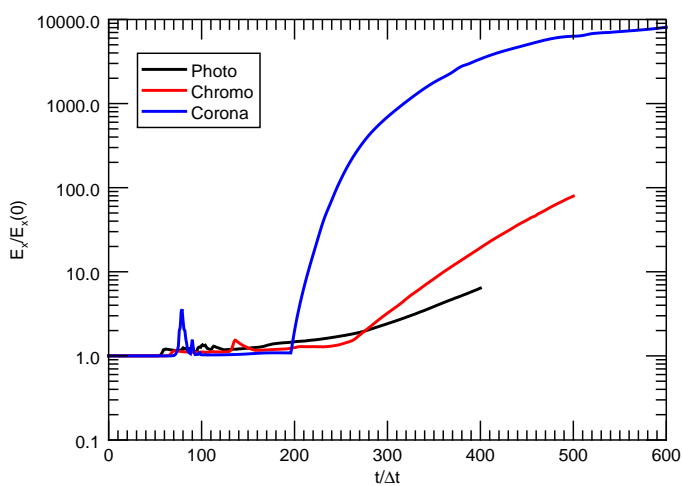}
\caption{Normalised electric field for the three atmospheric heights.}
\label{figcomefield}
\end{figure}

\begin{figure*}[!ht]
\begin{subfigure}{.48\textwidth}
(a)
\end{subfigure}
\begin{subfigure}{.48\textwidth}
(b)
\end{subfigure}\\
\centering
\begin{subfigure}{.48\textwidth}
\includegraphics[scale=0.55,clip=true, trim=2cm 7cm 2cm 8cm]{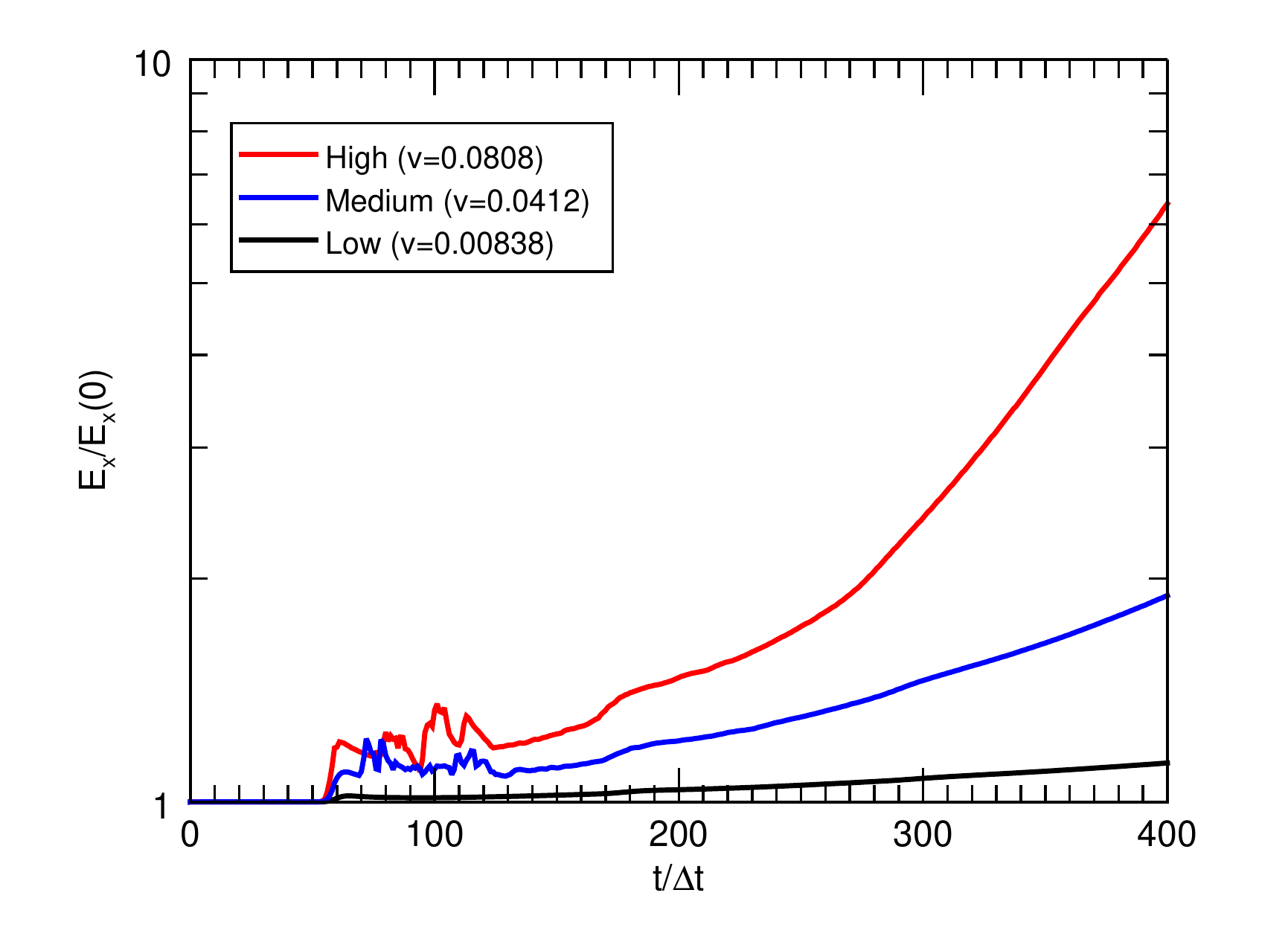}
\end{subfigure}
\begin{subfigure}{.48\textwidth}
\includegraphics[scale=0.55,clip=true, trim=2cm 7cm 2cm 8cm]{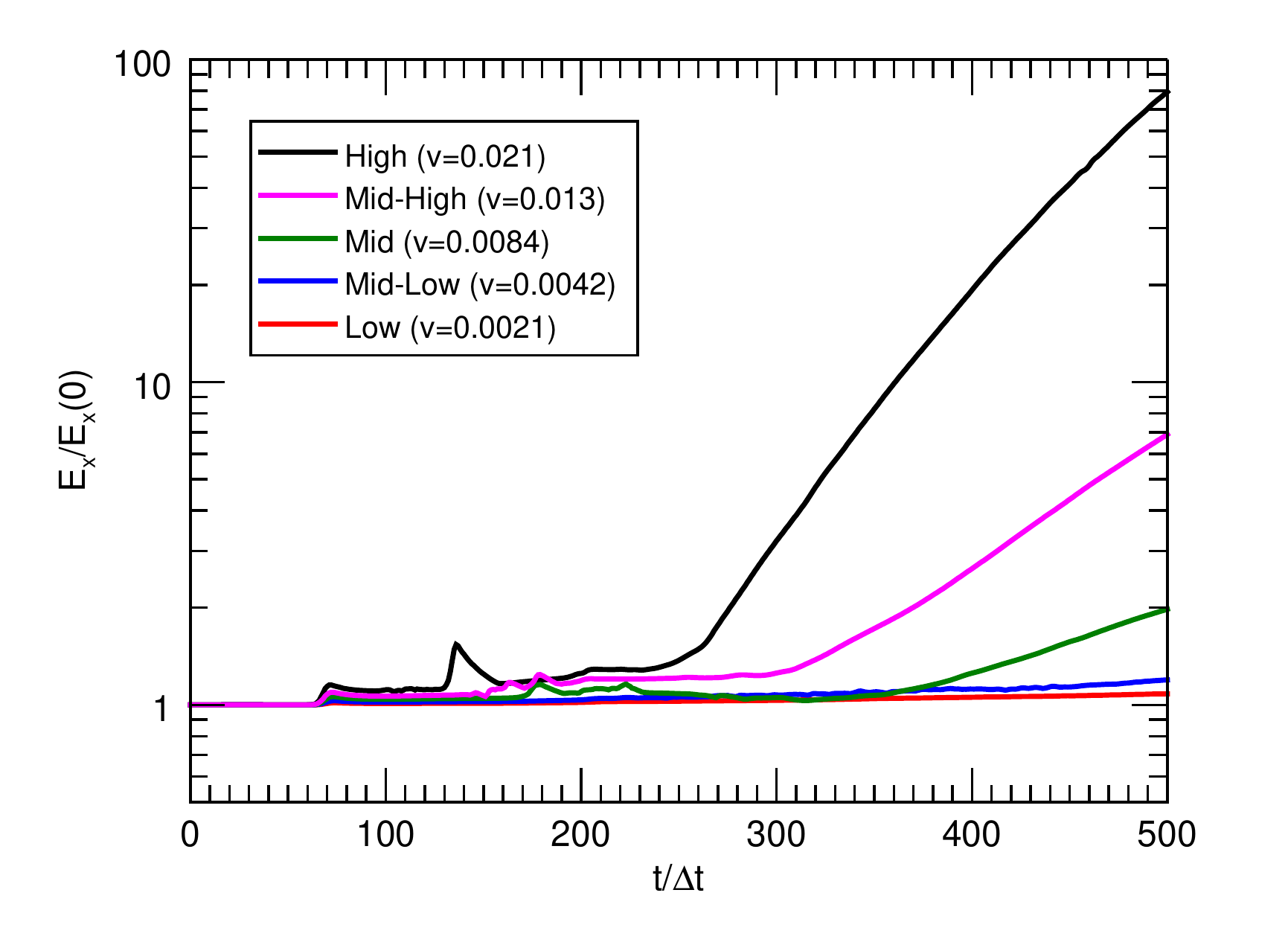}
\end{subfigure}
\caption{Normalised electric field in (a) the photospheric case and (b) the chromospheric case with different velocity drivers.}
\label{figcomphotoe}
\end{figure*}

\begin{figure}[!h]
\centering
\includegraphics[scale=0.55,clip=true, trim=3cm 7cm 2cm 8cm]{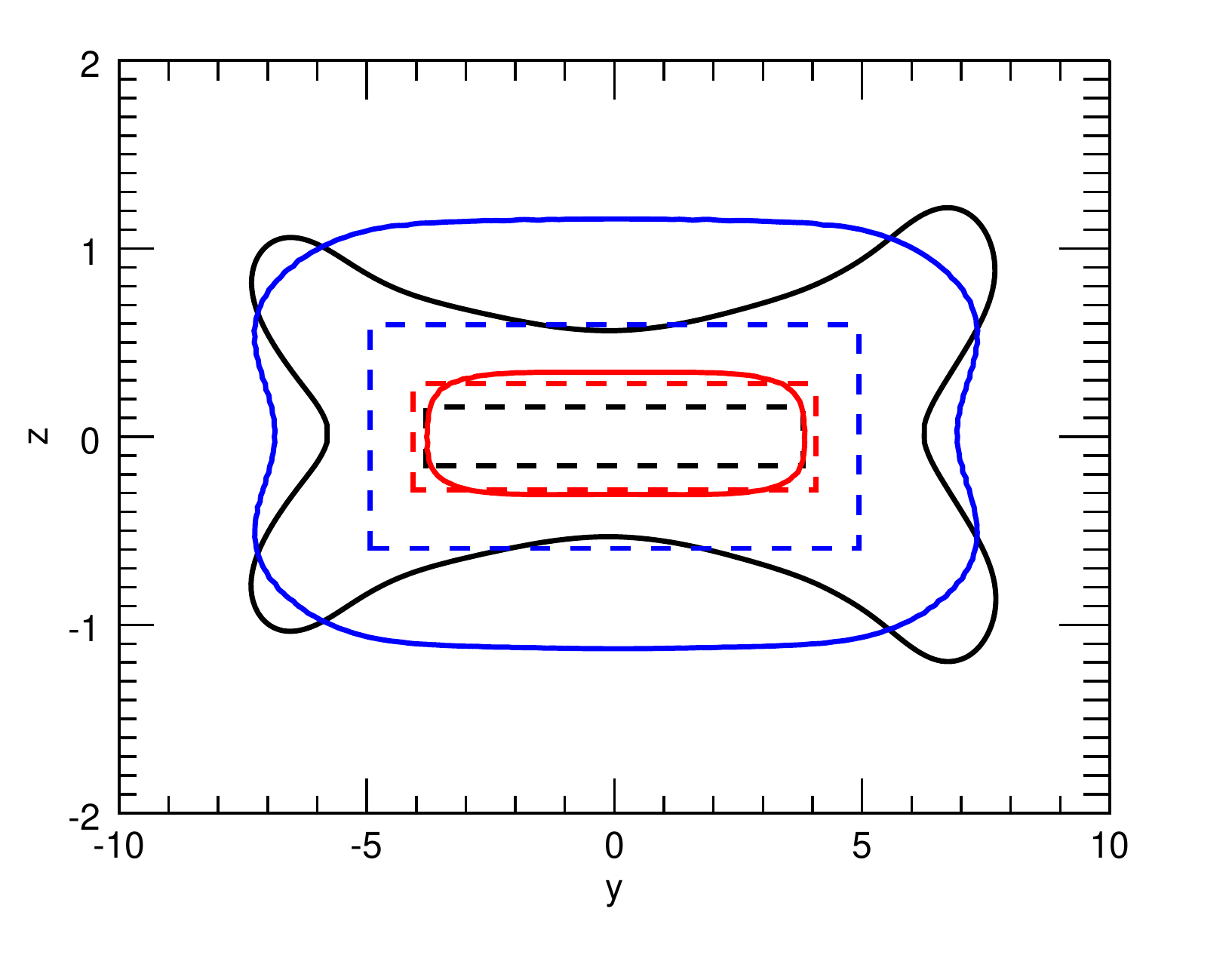}
\caption{Shape of the heating region for the photospheric (black), chromospheric (red) and coronal (blue). The level is half the maximum heating, i.e. a temperature increase of 1.5\% for photosphere and chromosphere and 2\% for the corona. The diffusion region of each case is overplotted with a dashed line.}
\label{figcomheat}
\end{figure}

\subsection{Heating}

There is a slight difference in the maximum percentage temperature increase when changing the neutral fraction. For a neutral fraction of $\xi _n =0.5$ and $\xi _n = 0.1$ the temperature in the centre of the domain increases by approximately 280K. For $\xi _n= 0.9$ there is temperature increase of approximately 410K. The perpendicular resistiviy depends on $\xi ^2$ (Equation (\ref{eqneta})) and hence the additional diffusion due to the $\eta _\perp J_\perp$ term is significantly higher when $\xi _n =0.9$ than when $\xi _n =0.5$ or $\xi _n =0.1$, producing more of a temperature increase. This implies that current sheets lower in the solar atmosphere (i.e. photosphere or lower chromosphere) produce more local heating. The distribution of the heated plasma is roughly identical in all cases and confined to the reconnection region.

\section{Onset of reconnection: photospheric vs chromospheric vs coronal environments} \label{sec_comp}


We can compare the onset of a reconnection event occurring at different atmospheric heights now that the different parameters have been investigated separately. The driver magnitude in all cases has been chosen such that the Alfv\'en Mach number on the inflow approaches but never exceeds unity. The key parameters in the different cases are in Table \ref{tabbasepar3}.

The output data for the simulations have been scaled such that the time for a fast-mode wave to propagate across the domain is approximately $100 \Delta t$. This means the wavefronts collide at time $t \approx 50 \Delta t$. The values of $\Delta t$ are 100, 38.5 and 24 seconds for photosphere, chromosphere and corona respectively. 

\begin{table}[!h]
\caption{Photospheric, chromospheric and coronal parameters}
     $$ 
         \begin{array}{p{0.29\linewidth}lll}
            \hline
            \noalign{\smallskip}
            Parameter [Units]     &  \mbox{Photosphere} & \mbox{Chromosphere} & \mbox{Corona} \\
            \noalign{\smallskip}
            \hline
            \noalign{\smallskip}
            Plasma-$\beta$ & 1   & 0.1  & 0.002  \\
            Temperature [K] & 6000  & 9200  & 10^6          \\
            Neutral fraction $\xi _n$ & \approx 1 & \approx 0.5 & \approx 0 \\
            Resistivity [$\Omega \mbox{m}$] & 10^{-3} & 10^{-3} & 10^{-6}\\
            Density normalisation [kg m$^{-3}$] & 10^{-14} & 10^{-14} & 10^{-16} \\
            Time $\Delta t$ [$s$] & 100 & 38.5 & 24 \\
            \noalign{\smallskip}
            \hline
         \end{array}
     $$ 
\label{tabbasepar3}
\end{table}

\subsection{Size and extent of the reconnection region}

The half-length $L$ and half-width $\delta$ of the reconnection region are shown in Figure \ref{figcomdif}. In the figure time is scaled by $\Delta t$ and the photospheric case appears to narrow at the fastest rate. However, by looking at the rates of the linear change in $\delta$ and using $\Delta t$ from Table \ref{tabbasepar3}, one can calculate the rate $\Delta \delta / \Delta t$, that is the speed at which the reconnection region is narrowing. These values are approximately 0.213, 0.4132 and 0.208 km s$^{-1}$ for the photosphere, chromosphere and corona respectively. Therefore the fastest rate of change is in the chromosphere. The perpendicular diffusion in the chromospheric case produced a sharp current structure and hence the half-width $\delta$ (calculated as the HWHM of the current density) narrows at a faster rate than the other atmospheres.

\subsection{Electric field evolution}

Figure \ref{figcomefield} shows the electric field in each case normalised by its electric field at time $t=0$. The three cases demonstrate very different behaviour during the advection-dominated phase. There are orders of magnitude difference between the maximum electric field in each case. In the photosphere, the electric field exhibits exponential behaviour (Figure \ref{figcomefield}) but has a very low gradient. We observe two separate exponential phases with different gradients: one between 180$\Delta t$ and 280$\Delta t$ corresponding to the tail end of the development phase, and one from 280$\Delta t$ and 400$\Delta t$ with a slightly steeper gradient that corresponds to the advection-dominated phase. The chromosphere has a very flat electric field until $t=270 \Delta t$ when the advection-dominated phase starts and the $\textbf{v} \times \textbf{B}$ term dominates. The corona has the highest electric field gradient, again corresponding to the $\textbf{v} \times \textbf{B}$ term dominating the electric field.

The electric field signature also depends on the driver velocity. This was analysed for the coronal case in Section \ref{sec_vel}. The electric field for different driver velocities in the photosphere and chromosphere are shown in Figures \ref{figcomphotoe}(a) and \ref{figcomphotoe}(b) respectively. The different driver velocities are shown in Table \ref{tabvelphoto} for the photospheric case and Table \ref{tabvelpartial} for the chromospheric case. In the photosphere, only the fastest driver yields a steep gradient in electric field at late times. This occurs when the advection and diffusion terms are of the same order. In the chromosphere, one can produce behaviour in the electric field similar to both photospheric and coronal cases depending on the magnitude of the velocity driver. The fastest driver has comparable behaviour to the coronal case: a steep gradient in electric field and advection dominates at late times. However for slow drivers there is very little change in the electric field and the diffusion dominates, similar to the slow photospheric case. Therefore reconnection occurring at chromospheric levels is highly dependent on the magnitude of the velocity driver. 

\begin{table}[!h]
\caption{Driver Alfv\'en Mach numbers for the photospheric case.}
     $$ 
         \begin{array}{p{0.5\linewidth}l}
            \hline
            \noalign{\smallskip}
            Driver name  &  \mbox{Alfv\'en Mach Number}  \\
            \noalign{\smallskip}
            \hline
            \noalign{\smallskip}
            High & 0.0808             \\
            Medium & 0.0412  \\
            Low & 0.00838     \\
            \noalign{\smallskip}
            \hline
         \end{array}
     $$ 
\label{tabvelphoto}
\end{table}

\begin{table}[!h]
\caption{Driver Alfv\'en Mach numbers for the chromospheric case.}
     $$ 
         \begin{array}{p{0.5\linewidth}l}
            \hline
            \noalign{\smallskip}
            Driver name  &  \mbox{Alfv\'en Mach Number}  \\
            \noalign{\smallskip}
            \hline
            \noalign{\smallskip}
            High & 0.021     \\
            Mid-High & 0.013 \\
            Mid & 0.0084 \\
            	Mid-Low & 0.0042  \\
            	Low & 0.0021             \\
            \noalign{\smallskip}
            \hline
         \end{array}
     $$ 
\label{tabvelpartial}
\end{table}

\subsection{Heating: change in internal energy}

The heating can be calculated in kelvin. In all cases the maximum heating occurs at the centre of the domain. The original values of the temperature are 6000, 9200 and $10^6$ K for the photosphere, chromosphere and corona (Table \ref{tabbasepar3}). The temperature change at the end of the simulation is approximately 186, 267 and 39456 K in the three cases. This corresponds to a 3.1\%, 2.9\% and 3.9\% rise in temperature over the simulation for the photosphere, chromosphere and corona. 

The shape of the heating region also varies in the different cases, see Figure \ref{figcomheat}. In all cases, the bulk of the heating is created in the diffusion region. For the photospheric case, the heating is a combination of two effects: Ohmic heating generated in the diffusion region is carried out along the double Y-shaped slow-mode shocks that exist on the interface between inflow and outflow regions, and some heat is created directly by the shock. This double Y-shape is similar to the shape of the outflow jet and termination shocks in \cite{Forbes1988}. The coronal case produces a fairly evenly distributed heating region with very little heat being carried out along the magnetic field lines. The chromospheric case has a very localised heating region of approximately the same size as the reconnection region. This is due to the power-law region in the current density creating a highly localised heating region.

Reconnection events generated in this setup deposit heat very differently in the surrounding plasma. In the coronal case, the heat is spread uniformly in a large area around the reconnection region. Chromospheric reconnection creates a heating signature that is entirely localised to the diffusion region. In the photosphere, the heat is distributed by shocks and flows around the reconnection site.

\section{Conclusions}

In this paper the onset of 2D magnetic reconnection has been investigated at different atmospheric heights of the Sun, namely the solar corona, chromosphere and photosphere, by using an external sub-Alfv\'enic velocity driver specified perpendicular to a Harris current sheet. This allows us to investigate the early behaviour of magnetic reconnection in a naturalistic manner. As waves and flows are ubiquitous in the solar atmosphere, these waves and flows can, at some time, encounter a current sheet and this can initiate reconnection as described in this paper. This is the physical interpretation of our velocity driver. Furthermore, by choosing different physical parameters, we can investigate the signatures of the reconnection onset as a result of this driver at different solar atmospheric layers, namely the photosphere, chromosphere and corona. 

The process has been separated into three phases: reflection, development and advection-dominated. When the velocity wavefront hits the centre of the equilibrium Harris current sheet there is a fast-mode reflection. This causes the current sheet to fluctuate in width. During the development phase, the current density starts to rise due to the inflow generated by the driver and magnetic field lines begin to reconnect. Finally during the advection-dominated phase there is clear acceleration of the plasma exiting the reconnection region. With reconnection initiated with this type of velocity driver, the reconnection appears to be fairly independent of the resistivity; the current density $J _x$ and outflow velocity $v_y$ remain identical when the resistivity is changed. Note that the specified resistivity is resolved in our grid (Appendix \ref{appa}). The fundamental reconnection structure is formed from the collision of the wavefronts. Acceleration of the plasma occurs, however the advection term in Ohm's law is dominant over the diffusion term. The rate at which magnetic flux can enter the diffusion region is determined by the velocity amplitude. This implies that for this configuration in the solar atmosphere, a large increase in electric field can only occur for large velocity amplitudes.

For the photospheric case, a large increase in electric field was only obtained as the velocity flowing into the reconnection region approached the Alfv\'en speed. The high gas pressure results in a very narrow current peak at the centre of the domain and hence a narrow reconnection region. This high gas pressure also means that more energy is required to accelerate the plasma, producing a weak electric field. The heat is carried out along slow-mode shocks that exist on the interface between the inflow and outflow regimes. The dense plasma limits the rate at which magnetic flux can enter the reconnection region during the onset of magnetic reconnection. The implication for the solar photosphere is that reconnection is less readily achieved, compared to other atmospheric layers, and requires a very large inflow amplitude.

In the corona, the electric field shows a far larger increase, compared to the other atmospheric layers. The advection term becomes four orders of magnitude larger than the diffusion term in Ohm's law at late times. The low plasma-$\beta$ means that plasma can easily be accelerated away from the reconnection region, resulting in a large velocity and hence a large advection term. 

The chromospheric case has atmospheric conditions somewhere between the corona and the photosphere. The components of Ohm's law behave much like the coronal case where the advection term is far larger than the diffusion term at late times. However the difference is approximately two orders of magnitude in the chromosphere, compared to four orders of magnitude in the corona. The inclusion of partial ionisation results in a power-law region in the distribution of the current on the inflow that was not present in the coronal or photospheric cases. However, the partial ionisation does not appear to have a significant effect on the outflow velocity or electric field signatures when using this driver. The initial onset of reconnection in the solar chromosphere behaves similarly to the fully-ionised coronal case, however the effects of partial ionisation become more important as the reconnection evolves.

There is a fundamental difference in the heating region in the three cases. The corona has a large regular-shaped heating region around the reconnection region. The low density in the corona allows the heat to spread fairly uniformly around the reconnection region. The photosphere appears to have heating that is guided along the double Y-shaped slow-mode shocks, surrounding the reconnection region. The chromospheric case has a very localised heating region almost the same size as the diffusion region. This is a result of the partial ionisation; the structure of the current density is a peak in the centre, surrounded by a power law region. This narrow peak in $J_x$ results in localised Ohmic heating and hence a heating region that is localised to the diffusion region. The heat generated by reconnection in the corona is distributed over a large area, whereas reconnection at chromospheric and photospheric levels produced localised heating as a signature of reconnection.

The reconnection event behaves Sweet-Parker-like in some ways, and Petschek-like in others. The reconnection region is narrow and elongated, as in Sweet-Parker. However weak slow-mode shocks form on the interface between the inflow and outflow and the outflow velocity is closer to the Petschek speed, where $v_{outflow} = v_{A(inflow)} \sqrt{\rho _{inflow} / \rho _{outflow}}$. The inflow magnetic field is stratified, thus the general behaviour is similar to the flux pile-up models \citep{Priest1986}.

In this paper, we consider the onset of 2D magnetic reconnection. A test was also performed in 2.5D using a constant magnetic field in the invariant $x-$direction. The main result of this was that the Alfv\'en speed at the centre of the domain does not reduce to zero, resulting in less flux pile-up on the inflow region. Qualitatively the results were the same between the 2D and 2.5D cases. Expanding to 3D opens up the potential for more complex configurations, however we expect that qualitatively the results would be comparable to this 2D study if they occur is a similar setup.

Reconnection is ubiquitous in the solar atmosphere and plays a key role in many phenomena occurring at photospheric, chromospheric and coronal levels. In this paper, numerical simulations were performed investigating the transient onset of magnetic reconnection at different atmospheric heights. This was achieved by using an external driver to trigger reconnection from an equilibrium Harris current sheet. The amount of flux that can enter the diffusion region is determined by the velocity of the inflow. A lower plasma-$\beta$ means a low gas pressure and hence magnetic field can be pushed together using lower velocities. This creates a larger change in the electric field signature of the reconnection event. This sharp rise in electric field is obtained for all tested velocity drivers. However the magnitude of this rise is dependent on the amplitude of the velocity driver; a higher amplitude driver produces a larger rise in electric field. 
The high plasma pressure (relative to the magnetic pressure) in the photospheric case leads to a narrower current sheet, compared to coronal and chromospheric cases.
For a high-$\beta$ plasma, a sharp rise in electric field is obtained only as the inflow into the reconnection region approaches the Alfv\'en speed. The rate at which magnetic flux enters the diffusion region is the key parameter determining the onset of magnetic reconnection.

\begin{acknowledgements}
The authors acknowledge IDL support provided by STFC.   A. Hillier is supported by his STFC Ernest Rutherford Fellowship grant number ST/L00397X/2. J.A. McLaughlin acknowledges generous support from the Leverhulme Trust and this work was funded by a Leverhulme Trust Research Project Grant: RPG-2015-075.

\end{acknowledgements}

\appendix

\section{Resistivity dependence} \label{appa}

\paragraph{Numerical resistivity:} The numerical resistivity on our grid was calculated by specifying a Harris current sheet  as the initial condition and running Lare3D specifying $\eta = 0$. The current sheet will collapse due to the numerical resistivity $\eta _n$. The numerical resistivity can then be calculated by solving the diffusion equation:
\begin{equation}
\frac{\partial \textbf{B}}{\partial t} = \eta _n \frac{\partial ^2 \textbf{B}}{\partial z^2}. \label{eqnetan}
\end{equation} 
Simulations are performed with two invariant directions and a grid density 512 cells over 32 Mm, corresponding to the spatial resolution used in the rest of the paper. This simulation was performed for a physical time of 20,000 seconds (corresponding to approximately 15,000 numerical iterations). The initial and final states are used to calculate $\eta _n$ from Equation(\ref{eqnetan}). The numerical resistivity calculated is of the order $\eta _n =10^{-8}$ $\Omega$m.

\paragraph{Order of magnitude argument:} The calculated value of numerical resistivity is very small ($\mathcal{O}(-8)$). An order of magnitude analysis can be used to compare this to previous work. The resistivity (using normalisation values) is effectively:
\begin{equation}
\eta = l_0 v_0 \mu _0 = l_0 \mu _0 \frac{B_0}{\sqrt{\mu _0 \rho _0}}.
\end{equation} 
Coronal simulations in the literature that use Lare3d typically use a length scale of $10^6$ m and have a resistivity of order $-3$ to $-4$  $\Omega$m. Here we are interested in localised reconnection so our length scale is smaller, $l_0 =10^4$ m. This effectively means that we can use a resistivity that is of order $-5$ to $-6$ $\Omega$m without the output becoming dominated by numerical resistivity.

\paragraph{Physical resistivity:} The coronal simulation was repeated using a specified resistivity of $\eta = 10^{-3}$ $\Omega$m. (The coronal resistivity used elsewhere in this paper is $10^{-6}$ $\Omega$m.) The only variable this changes is $B_z$ on the outflow plane: it becomes larger as the simulation advances. This means that there is a slight change in behaviour when the resistivity is increased, further validating our calculation of numerical resistivity. The variable $B_z$ on the outflow region is significantly smaller, by several orders of magnitude, than the inflow $B_y$. As a result, the change in $B_z$, whilst validating our choice of resistivity, does not correspond to changes one may expect from changing the resistivity. The reconnection would have to become significantly more developed in order to notice changes in other parameters from changing the resistivity. This further suggests that the onset of reconnection is dominated by advection effects. 

\begin{figure}[!h]
\centering
\includegraphics[scale=0.55,clip=true, trim=2cm 7cm 2cm 8cm]{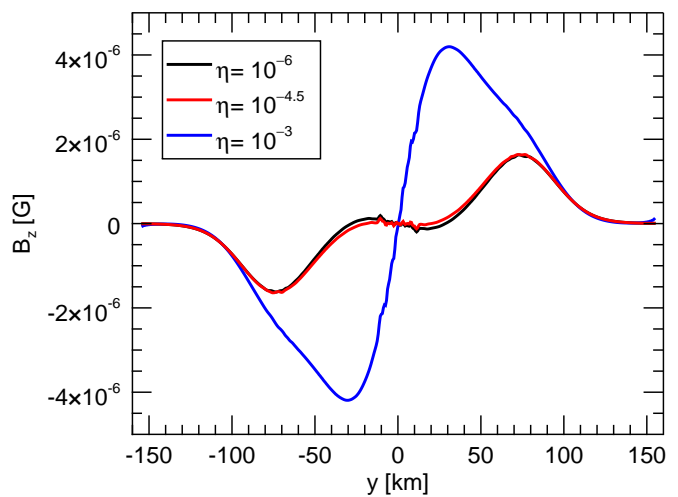}
\caption{Reconnected magnetic field $B_z$ along the centre line at time $t=600\Delta t$ for different values of resistivity. The resistivity $\eta$ is in units $\Omega$m.}
\label{recon_mag_field}
\end{figure}

\paragraph{$B_z$ scaling with resistivity:} Figure \ref{recon_mag_field} gives the reconnected magnetic field ($B_z$) along the $z=0$ line at the time $t=600\Delta t$ for three different resistivities ($\eta=10^{-3}$, $10^{-4.5}$ and $10^{-6}$ $\Omega$m). It is clear from this figure that the amount of reconnected flux is different for each resistivity value. It is possible to create a simple model to estimate the change in the reconnected flux. Firstly we assume that the current is dominated by the $B_y$ field component, so we set $J_x\approx -\partial B_y/\partial z$. As the early stages of reconnection have weak flows, then we can estimate the temporal evolution of $B_z$ is dominated by diffusion, i.e. $\partial B_z/\partial t\sim -\eta\partial J_x/\partial y$. If $J_x$ remains relatively constant (and determined by the inflow magnitude and structure) once the reconnection has started, then $B_z(t)\sim -\eta t \partial J_x/\partial y$. Using this we can estimate the change in magnitude of the reconnected flux ($\epsilon$) for a given change in resistivity as:
\begin{equation}
\epsilon=\int_{-Y}^{Y}|B_{z,\eta_1}-B_{z,\eta_2}|\partial x\sim \eta_2\int_{-Y}^{Y}\left|t \frac{\partial J_x}{\partial y}\left(\frac{\eta_1}{\eta_2}-1\right)\right|\partial x.
\end{equation}
From this we would expect that if we calculate the value of $\epsilon$ for $\eta_1=10^{-3}$ $\Omega$m and $\eta_2=10^{-4.5}$ $\Omega$m, or $\eta_1=10^{-4.5}$ $\Omega$m and $\eta_2=10^{-6}$ $\Omega$m we would get a difference of approximately $10^{1.5}\sim 30$ between them. For the simulations shown in Figure \ref{recon_mag_field} with $\eta_1=10^{-3}$ $\Omega$m and $\eta_2=10^{-4.5}$ $\Omega$m we get $\epsilon =6.5\times 10^{-5}$, and with $\eta_1=10^{-4.5}$ $\Omega$m and $\eta_2=10^{-6}$ $\Omega$m we get $\epsilon =1.9\times 10^{-6}$ which differ by a factor of $\sim 30$ as we predict from our simple model. This test does imply that we are able to see differences in the reconnection behaviour down to very small resistivity values, highlighting that we are not in a regime dominated by numerical resistivity for the simulations shown in this paper. We would like to note that we are not claiming that this would hold into the fully developed stage of reconnection, because in the later stages of reconnection when the current sheet dynamically thins the smaller the resistivity value the sooner the system will become dominated by numerical resistivity.

%
%

   \bibliographystyle{aa} 
   \bibliography{snowbib.bib} 

\begin{thebibliography}{52}
\expandafter\ifx\csname natexlab\endcsname\relax\def\natexlab#1{#1}\fi

\bibitem[{Arber {et~al.}(2001)Arber, Longbottom, Gerrard, \& Milne}]{Arber2001}
Arber, T., Longbottom, A.~W., Gerrard, C., \& Milne, A.~M. 2001, J. Comput.
  Phys, 171

\bibitem[{{Arber} {et~al.}(2009){Arber}, {Botha}, \& {Brady}}]{Arber2009}
{Arber}, T.~D., {Botha}, G.~J.~J., \& {Brady}, C.~S. 2009, \apj, 705, 1183

\bibitem[{{Arber} \& {Haynes}(2006)}]{Arber2006}
{Arber}, T.~D. \& {Haynes}, M. 2006, Physics of Plasmas, 13, 112105

\bibitem[{{Arber} {et~al.}(2007){Arber}, {Haynes}, \& {Leake}}]{Arber2007}
{Arber}, T.~D., {Haynes}, M., \& {Leake}, J.~E. 2007, \apj, 666, 541

\bibitem[{{Aschwanden}(2005)}]{Aschwanden2005}
{Aschwanden}, M.~J. 2005, {Physics of the Solar Corona. An Introduction with
  Problems and Solutions (2nd edition)}

\bibitem[{{Athay} \& {Thomas}(1961)}]{Thomas1961}
{Athay}, R.~G. \& {Thomas}, R.~N. 1961, {Physics of the solar chromosphere}

\bibitem[{{Bhattacharjee}(2004)}]{Bhattacharjee2004}
{Bhattacharjee}, A. 2004, \araa, 42, 365

\bibitem[{{Brandenburg} \& {Zweibel}(1994)}]{Brandenburg1994}
{Brandenburg}, A. \& {Zweibel}, E.~G. 1994, \apjl, 427, L91

\bibitem[{{De Moortel}(2005)}]{2005RSPTA.363.2743D}
{De Moortel}, I. 2005, Philosophical Transactions of the Royal Society of
  London Series A, 363, 2743

\bibitem[{Dewar {et~al.}(2013)Dewar, Bhattacharjee, Kulsrud, \&
  Wright}]{Dewar2013}
Dewar, R.~L., Bhattacharjee, A., Kulsrud, R.~M., \& Wright, A.~M. 2013, Physics
  of Plasmas, 20, 082103

\bibitem[{{Edwin} \& {Roberts}(1983)}]{1983SoPh...88..179E}
{Edwin}, P.~M. \& {Roberts}, B. 1983, \solphys, 88, 179

\bibitem[{{Ellerman}(1917)}]{Ellerman1917}
{Ellerman}, F. 1917, \apj, 46, 298

\bibitem[{{Fitzpatrick}(2003)}]{Fitzpatrick2003b}
{Fitzpatrick}, R. 2003, Physics of Plasmas, 10, 2304

\bibitem[{{Fitzpatrick} {et~al.}(2003){Fitzpatrick}, {Bhattacharjee}, {Ma}, \&
  {Linde}}]{Fitzpatrick2003}
{Fitzpatrick}, R., {Bhattacharjee}, A., {Ma}, Z.-W., \& {Linde}, T. 2003,
  Physics of Plasmas, 10, 4284

\bibitem[{{Forbes}(1988)}]{Forbes1988}
{Forbes}, T.~G. 1988, \solphys, 117, 97

\bibitem[{{Hahm} \& {Kulsrud}(1985)}]{Hahm1985}
{Hahm}, T.~S. \& {Kulsrud}, R.~M. 1985, Physics of Fluids, 28, 2412

\bibitem[{{Hillier} {et~al.}(2016){Hillier}, {Takasao}, \&
  {Nakamura}}]{Hillier2016}
{Hillier}, A., {Takasao}, S., \& {Nakamura}, N. 2016, \aap, 591, A112

\bibitem[{{Jess} {et~al.}(2015){Jess}, {Morton}, {Verth}, {Fedun}, {Grant}, \&
  {Giagkiozis}}]{Jess2015}
{Jess}, D.~B., {Morton}, R.~J., {Verth}, G., {et~al.} 2015, \ssr, 190, 103

\bibitem[{{Katsukawa} {et~al.}(2007){Katsukawa}, {Berger}, {Ichimoto}, {Lites},
  {Nagata}, {Shimizu}, {Shine}, {Suematsu}, {Tarbell}, {Title}, \&
  {Tsuneta}}]{Katsukawa2007}
{Katsukawa}, Y., {Berger}, T.~E., {Ichimoto}, K., {et~al.} 2007, Science, 318,
  1594

\bibitem[{Leake {et~al.}(2005)Leake, Arber, \& Khodachenko}]{Leake2005}
Leake, J.~E., Arber, T., \& Khodachenko, M. 2005, Astronomy \& Astrophysics,
  442, 1091

\bibitem[{{Leake} \& {Arber}(2006)}]{Leake2006}
{Leake}, J.~E. \& {Arber}, T.~D. 2006, A\&~A, 450, 805

\bibitem[{{Leake} {et~al.}(2012){Leake}, {Lukin}, {Linton}, \&
  {Meier}}]{Leake2012}
{Leake}, J.~E., {Lukin}, V.~S., {Linton}, M.~G., \& {Meier}, E.~T. 2012, ApJ,
  760, 109

\bibitem[{{Malyshkin} \& {Kulsrud}(2010)}]{Malyshkin2010}
{Malyshkin}, L.~M. \& {Kulsrud}, R.~M. 2010, Physica Scripta Volume T, 142,
  014034

\bibitem[{{Malyshkin} \& {Zweibel}(2011)}]{Malyshkin2011}
{Malyshkin}, L.~M. \& {Zweibel}, E.~G. 2011, \apj, 739, 72

\bibitem[{{Moore} {et~al.}(2001){Moore}, {Sterling}, {Hudson}, \&
  {Lemen}}]{Moore2001}
{Moore}, R.~L., {Sterling}, A.~C., {Hudson}, H.~S., \& {Lemen}, J.~R. 2001,
  \apj, 552, 833

\bibitem[{{Morita} {et~al.}(2010){Morita}, {Shibata}, {Ueno}, {Ichimoto},
  {Kitai}, \& {Otsuji}}]{Morita2010}
{Morita}, S., {Shibata}, K., {Ueno}, S., {et~al.} 2010, \pasj, 62, 901

\bibitem[{{Nakariakov}(2007)}]{2007AdSpR..39.1804N}
{Nakariakov}, V.~M. 2007, Advances in Space Research, 39, 1804

\bibitem[{{Nakariakov} \& {Verwichte}(2005)}]{2005LRSP....2....3N}
{Nakariakov}, V.~M. \& {Verwichte}, E. 2005, Living Reviews in Solar Physics,
  2, 3

\bibitem[{{Nelson} {et~al.}(2016){Nelson}, {Doyle}, \&
  {Erd{\'e}lyi}}]{Nelson2016}
{Nelson}, C.~J., {Doyle}, J.~G., \& {Erd{\'e}lyi}, R. 2016, \mnras, 463, 2190

\bibitem[{Parker(1957)}]{Parker1957}
Parker, E.~N. 1957, Journal of Geophysical Research, 62, 509

\bibitem[{{Petschek}(1964)}]{Petschek1964}
{Petschek}, H.~E. 1964, NASA Special Publication, 50, 425

\bibitem[{{Pontin}(2012)}]{Pontin2012}
{Pontin}, D.~I. 2012, Philosophical Transactions of the Royal Society of London
  Series A, 370, 3169

\bibitem[{Priest(1984)}]{priest1984solar}
Priest, E. 1984, Solar Magnetohydrodynamics, Geophysics and Astrophysics
  Monographs (Springer Netherlands)

\bibitem[{{Priest} \& {Forbes}(2000)}]{Priest2000}
{Priest}, E. \& {Forbes}, T. 2000, {Magnetic Reconnection}, 612

\bibitem[{{Priest} \& {Forbes}(1986)}]{Priest1986}
{Priest}, E.~R. \& {Forbes}, T.~G. 1986, J.G.R., 91, 5579

\bibitem[{{Reid} {et~al.}(2016){Reid}, {Mathioudakis}, {Doyle}, {Scullion},
  {Nelson}, {Henriques}, \& {Ray}}]{Reid2016}
{Reid}, A., {Mathioudakis}, M., {Doyle}, J.~G., {et~al.} 2016, \apj, 823, 110

\bibitem[{{Sakai} \& {Smith}(2008)}]{Sakai2008}
{Sakai}, J.~I. \& {Smith}, P.~D. 2008, \apjl, 687, L127

\bibitem[{{Savcheva} {et~al.}(2007){Savcheva}, {Cirtain}, {Deluca},
  {Lundquist}, {Golub}, {Weber}, {Shimojo}, {Shibasaki}, {Sakao}, {Narukage},
  {Tsuneta}, \& {Kano}}]{Savcheva2007}
{Savcheva}, A., {Cirtain}, J., {Deluca}, E.~E., {et~al.} 2007, \pasj, 59, S771

\bibitem[{{Shibata} {et~al.}(1992){Shibata}, {Ishido}, {Acton}, {Strong},
  {Hirayama}, {Uchida}, {McAllister}, {Matsumoto}, {Tsuneta}, {Shimizu},
  {Hara}, {Sakurai}, {Ichimoto}, {Nishino}, \& {Ogawara}}]{Shibata1992}
{Shibata}, K., {Ishido}, Y., {Acton}, L.~W., {et~al.} 1992, \pasj, 44, L173

\bibitem[{{Shibata} {et~al.}(2007){Shibata}, {Nakamura}, {Matsumoto}, {Otsuji},
  {Okamoto}, {Nishizuka}, {Kawate}, {Watanabe}, {Nagata}, {UeNo}, {Kitai},
  {Nozawa}, {Tsuneta}, {Suematsu}, {Ichimoto}, {Shimizu}, {Katsukawa},
  {Tarbell}, {Berger}, {Lites}, {Shine}, \& {Title}}]{Shibata2007}
{Shibata}, K., {Nakamura}, T., {Matsumoto}, T., {et~al.} 2007, Science, 318,
  1591

\bibitem[{{Singh} {et~al.}(2015){Singh}, {Hillier}, {Isobe}, \&
  {Shibata}}]{Singh2015}
{Singh}, K.~A.~P., {Hillier}, A., {Isobe}, H., \& {Shibata}, K. 2015, \pasj,
  67, 96

\bibitem[{{Smith} \& {Sakai}(2008)}]{Smith2008}
{Smith}, P.~D. \& {Sakai}, J.~I. 2008, \aap, 486, 569

\bibitem[{Sweet(1958)}]{Sweet1958}
Sweet, P.~A. 1958, in Electromagnetic phenomena in cosmical physics, Vol.~6,
  123

\bibitem[{{Tomczyk} {et~al.}(2007){Tomczyk}, {McIntosh}, {Keil}, {Judge},
  {Schad}, {Seeley}, \& {Edmondson}}]{2007Sci...317.1192T}
{Tomczyk}, S., {McIntosh}, S.~W., {Keil}, S.~L., {et~al.} 2007, Science, 317,
  1192

\bibitem[{{Ugai} \& {Tsuda}(1977)}]{Ugai1977}
{Ugai}, M. \& {Tsuda}, T. 1977, Journal of Plasma Physics, 17, 337

\bibitem[{{Wang} \& {Bhattacharjee}(1992)}]{Wang1992}
{Wang}, X. \& {Bhattacharjee}, A. 1992, Physics of Fluids B, 4, 1795

\bibitem[{{Wang} {et~al.}(1996){Wang}, {Ma}, \& {Bhattacharjee}}]{Wang1996}
{Wang}, X., {Ma}, Z.~W., \& {Bhattacharjee}, A. 1996, Physics of Plasmas, 3,
  2129

\bibitem[{{Yamada} {et~al.}(2010){Yamada}, {Kulsrud}, \& {Ji}}]{Yamada2010}
{Yamada}, M., {Kulsrud}, R., \& {Ji}, H. 2010, Reviews of Modern Physics, 82,
  603

\bibitem[{{Yamada} {et~al.}(2006){Yamada}, {Ren}, {Ji}, {Breslau}, {Gerhardt},
  {Kulsrud}, \& {Kuritsyn}}]{Yamada2006}
{Yamada}, M., {Ren}, Y., {Ji}, H., {et~al.} 2006, Physics of Plasmas, 13,
  052119

\bibitem[{{Zweibel}(1989)}]{Zweibel1989}
{Zweibel}, E.~G. 1989, \apj, 340, 550

\bibitem[{{Zweibel} {et~al.}(2011){Zweibel}, {Lawrence}, {Yoo}, {Ji}, {Yamada},
  \& {Malyshkin}}]{Zweibel2011}
{Zweibel}, E.~G., {Lawrence}, E., {Yoo}, J., {et~al.} 2011, Physics of Plasmas,
  18, 111211

\bibitem[{{Zweibel} \& {Yamada}(2009)}]{Zweibel2009}
{Zweibel}, E.~G. \& {Yamada}, M. 2009, \araa, 47, 291

\end{thebibliography}

\end{document}